\documentclass[12pt, onecolumn, amsmath, amssymb, a4paper]{aastex631}
\usepackage{times}
\usepackage{amsmath, bm}
\usepackage{graphicx}
\usepackage{wrapfig}
\usepackage{verbatim}
\usepackage{tikz}
\usepackage{color}

\usepackage{enumitem}

\usepackage[utf8]{inputenc}
\usepackage{newunicodechar}
\usepackage{tcolorbox}
\newunicodechar{−}{-}

\usepackage[T1]{fontenc}
\usepackage{aecompl}
\usepackage{calligra}
\DeclareMathAlphabet{\mathcalligra}{T1}{calligra}{m}{n}
\DeclareFontShape{T1}{calligra}{m}{n}{<->s*[2.2]callig15}{}

\usepackage{hyperref}
\hypersetup{
    pdfnewwindow=true,      
    colorlinks=true,       
    linkcolor=cyan,          
    citecolor=cyan,        
    filecolor=cyan,      
    urlcolor=cyan           
}

\def\apjl{ApJL}               
\def\apjs{ApJS}               
\def\aap{A\&A}                

\DeclareMathAlphabet{\mathcalligra}{T1}{calligra}{m}{n}
\DeclareFontShape{T1}{calligra}{m}{n}{<->s*[2.2]callig15}{}

\def\eg{\textit{e.g.}}
\def\ie{\textit{i.e.}}

\def\Msun{{M_{\odot}}}

\def\Dam{\mathcal{D}}
\def\Mdot{\dot{M}}

\def\Src{\mathcal{S}}
\def\Edd{\rm{Edd}}
\def\GW{\mathcal{GW}}
\def\res{\mathrm{res}}
\def\bin{\mathrm{bin}}
\def\orb{\mathrm{orb}}

\def\Vbet{\boldsymbol{\beta}}
\def\hatn{\mathbf{\hat{n}}}

\begin{document}
\title[]{Observational Signatures of Supermassive Black Hole Binaries}
\author[0000-0002-1271-6247]{Daniel J. D'Orazio}
\affiliation{Niels Bohr International Academy, Niels Bohr Institute, Blegdamsvej 17, 2100 Copenhagen, Denmark}
\email{daniel.dorazio@nbi.ku.dk}
\author[0000-0003-3579-2522]{Maria Charisi}
\affiliation{Department of Physics \& Astronomy, Vanderbilt University, 2301 Vanderbilt Place, Nashville, TN 37235, USA}
\email{maria.charisi@nanograv.org}

\begin{abstract} 
Despite solid theoretical and observational grounds for the pairing of supermassive black holes (SMBHs) after galaxy mergers, definitive evidence for the existence of close (sub-parsec) separation SMBH binaries (SMBHBs) approaching merger is yet to be found. This chapter reviews techniques aimed at discovering such SMBHBs in galactic nuclei. We motivate the search with a brief overview of SMBHB formation and evolution, and the gaps in our present-day theoretical understanding. We then present existing observational evidence for SMBHBs and discuss ongoing efforts to provide definitive evidence for a population at sub-parsec orbital separations, where many of the aforementioned theoretical gaps lie. We conclude with future prospects for discovery with electromagnetic (primarily time-domain) surveys, high-resolution imaging experiments, and low-frequency gravitational-wave detectors. 
\end{abstract}

\section{\large  Introduction}

In the 1960's the discovery of quasars as extra-galactic entities \citep{Schmidt:1963} and identification of their power source with accreting supermassive black holes \citep[SMBHs;][]{Robinson+1965, LyndenBell:1969} led towards the modern picture of SMBHs occupying the centers of nearly all massive galaxies \citep{KormendyRichstone:1995}. Soon after these first identifications of SMBHs as quasar power sources, \cite{ThorneBrag:1976} demonstrated that the initial collapse or mergers of SMBHs would generate bursts of gravitational radiation detectable at Earth. \cite{BBR:1980} carried this further by painting a first picture of the relevant astrophysical processes that could bring two single SMBHs together after collisions of galaxies. 
They additionally posited multiple signatures that would provide evidence for SMBH binaries (SMBHBs) and their pairings in galactic nuclei. They argued that the binary hardening process could create observable light deficits in the cores of galaxies, and that the orbital motion of SMBH pairs could generate peculiar radio jet morphology, emission line offsets, or photometric variability \citep[see also][]{Komberg:1968}.

In the forty years since the seminal work by \cite{BBR:1980}, the modeling of environmental interactions that do, or do not bring SMBHs towards merger is still an active topic at the heart of a major open problem in astrophysics: the final-parsec problem \citep[\S\ref{S:IIphys_procs};][]{Milosavljevic:2003:FPcP, MerrittMilos:2005:LRR,Colpi:2014}. While theoretical work has since developed solutions to the initial posing of the final parsec problem, \ie, identifying processes which will bring two SMBHs close enough to merge in a Hubble time (see, \eg, \S\ref{S:IIphys_procs}), the mystery of which processes operate in nature remains. In essence, this problem is similar to the current investigation of astrophysical formation channels of the stellar mass black hole binaries detected in gravitational waves (GWs) by LIGO-Virgo-KAGRA: there exist formation channels which could work in theory, but which of these operate in nature, and in what proportion is still unknown \citep[\eg,][]{Zevin_LIGOchans+2021}.
For the SMBHs, a concise, modern statement of the final parsec problem, or what we might now call the \textit{supermassive-merger problem}, is {\em galaxies containing SMBHs merge, but do their SMBHs also merge, and how?}

Further modeling of SMBHB environmental interactions is needed for reliable population predictions. But importantly, discovery of members of the population at different stages of the SMBHB lifespan is needed to test and hone the understanding of astrophysics that goes into these predictions. 
This review focuses on the latter, with a brief introduction to the former. We focus primarily on electromagnetic discovery for a few reasons:
\begin{enumerate}
    \item While GW observations of SMBHB inspiral and merger will give definitive evidence for the population at its end stages, electromagnetic identification offers a demographic probe at earlier stages that is critical for understanding the supermassive-merger problem, and the whole lifespan of the SMBHB (see Figure~\ref{Fig:BHBDemo}).
    \item Electromagnetic identification methods are varied and evolving, and it is unclear which (combination of) methods are the most promising for providing definitive evidence for SMBHBs. Hence, we attempt to make clear the ideas behind the different methods and give the pros and cons of each throughout.
    \item We are entering an era where unprecedented quantities of electromagnetic data will soon be available via photometric and spectroscopic surveys, \eg, the Vera Rubin Observatory's Legacy Survey of Space and Time \citep[LSST,][]{LSST:2019} and SDSS-V \citep{SDSSV+2017}, and surveys with the planned Roman Space Telescope \citep{RomanHLspecSurvey}; as well as high resolution radio and optical telescopes, \eg, the  Next Generation Very Large Array \citep[ngVLA,][]{BurkeSpolaor_nGVLA_SMBHBs:2018}, the Event Horizon Telescope \citep[EHT,][]{EHT_M87_I:2019} and its extensions \citep[\eg, ngEHT, EHE,][]{ngEHT_MMScience:2023, EHE:2022}, and GRAVITY+ \citep{GRAVITY+2023}. These, along with the James Webb Space Telescope \citep[JWST,][]{JWST+2006}, and  planned or proposed missions at multiple wavelengths, \eg, the ATHENA X-ray mission \citep{ATHENA:2013, ATHENA_AGN:2020} will provide new observational handles on the SMBHBs and guide methods for discovery.
\end{enumerate}

In \S\ref{S:IIphys_procs}, we briefly review the physical processes thought to bring SMBHBs together and provide some basic expectations for the SMBHB population. These expectations motivate where and how to look for SMBHBs at close separation, which we discuss in the opening of \S\ref{S:IIIObs}. The remainder of \S\ref{S:IIIObs} details SMBHB identification methods, quintessential candidates, if any, and the pros and cons of each method. We discuss future prospects for observational identification of SMBHBs in \S\ref{S:IVFuture}.
A non-complete, but extensive list of SMBHB candidates is compiled at the end of the chapter, in \S\ref{Sec:obs_summary}. 
We have not aimed for completeness in this review, but rather, have given a broad overview and introduction to promising detection methods while providing further focus on a few of the most exciting topics for us at this time. For other recent and related reviews see \citet{BurkeSpolaor_nGVLA_SMBHBs:2018, WangLi_ObsSBHBH_Rev:2020, Bogdanovic_EMSBHB_review+2022} and \S2 of \citet{LISAWP:2023}.


\section{\large  A Brief Primer on the SMBHB Lifespan }
\label{S:IIphys_procs}

To find something, you need to know where to look and what to look for. To gain insight we first review the SMBHB lifespan: the physical components as well as temporal and spatial scales associated with bringing two SMBHs together from galactic merger to the final SMBHB coalescence. A more in-depth look at these processes is provided in Chapter \textcolor{red}{3} of this book. 

\subsection{\large  Overview}
For our purposes, the SMBHB merger process can be described by an initial condition and three steps, which qualitatively resemble the picture put forward by \citet{BBR:1980}. 
        \paragraph{Initial Condition}
        The SMBH occupation fraction and galaxy merger rate at a given redshift provide the initial pairings of SMBHs at that redshift that may eventually form tight binaries at lower redshifts. While we have observational constraints from galaxy merger rates and the occurrence rates of dual active galactic nuclei (AGN, see \S\ref{SubS:Dual}), 
        it is still a challenge to 
        quantify the SMBH pairing rate and distribution across redshift \citep[\eg,][and references therein]{Koss_DUALAGN_WP+2019}.
        \paragraph{Large-Scale Evolution} The SMBHs delivered through galactic merger sink to the center of the newly formed galaxy via dynamical friction. After of order a galactic dynamical time ($\sim10^8$~yr), the two become bound relative to the surrounding stellar cluster, forming an SMBHB at $\mathcal{O}(1-10)$~pc separations. This limit is denoted by the dot-dashed grey line in Figure~\ref{Fig:BHBDemo} indicating where the binary orbital velocity is of order the surrounding velocity dispersion of stars \citep[the so-called ``hard-binary'' limit][]{BBR:1980, Milosavljevic:2003:FPcP}. Uncertainties in the efficacy of dynamical friction leading to binary formation at $\sim$kpc scales also hinder our current theoretical grasp on the SMBHB population \citep[\eg, see \S 2 of][and references therein]{LISAWP:2023}, but we do not focus on them here. 
        \paragraph{Small Scale Evolution} Gravitational radiation carries away energy and angular momentum from the binary which decreases its semi-major axis and circularizes the orbit. 
        A criterion for where GWs become important for binary evolution can be estimated to be where the binary lifetime due to gravitational radiation becomes shorter than the age of the universe.
        This happens when the separation, $a_{\GW}$, and the orbital period, $P_{\GW}$ of the binary are given by
    \begin{eqnarray}
        a_{\GW} &\approx& 0.049 \ \mathrm{pc} \ \left(\frac{q/(1+q)^2}{0.25}\right)^{1/4} \left( \frac{T}{1.3 \times 10^{10} \mathrm{yr} }\right)^{1/4} \left( \frac{M}{10^8 \Msun}\right)^{3/4}
        \nonumber \\ 
        P_{\GW} &\approx& 100.5 \ \mathrm{yr} \ \left(\frac{q/(1+q)^2}{0.25}\right)^{3/8} \left( \frac{T}{1.3 \times 10^{10} \mathrm{yr} }\right)^{3/8} \left( \frac{M}{10^8 \Msun}\right)^{5/8},
    \end{eqnarray}
        where $q=M_2/M_1\leq1$ is the mass ratio of the binary with total mass $M=M_1+M_2$ and $T$ is the GW-driven merger time.
        Only binaries with orbital separations smaller (or periods shorter) than this can be driven to merger by GWs alone. 
        There are other meaningful criteria for when binary evolution becomes dominated by GWs, but the above equations provide a useful reference which is agnostic of other processes operating near merger, \eg, gas decoupling \citep[][]{HKM09}.
        The steep dependence of the GW-driven merger time $T$ on the orbital separation $a$ (i.e. $T\propto a^4$) is in part responsible for the difficulty in driving the SMBHB through intermediate scale separations and the ``final parsec'' of its evolution discussed next.
        \paragraph{Intermediate Scale Evolution (sub-parsec)} At small enough binary separations, dynamical friction becomes inefficient, and binary evolution is driven by interactions with its immediate environment (\eg, stars). However, the number of stars on centrophilic orbits, which can remove energy and angular momentum from the binary, is quickly depleted. Initial treatments of the problem found that, if the binary interacts only with a spherical distribution of surrounding stars, the binary stalls in the final parsec before merger \citep{Milosavljevic:2003:FPcP}. This uncertainty in how SMBHBs might cross the final parsec on the path towards GW-driven merger is often referred to as the ``final-parsec problem,'' as discussed in the introduction to this Chapter and explored further in Chapter \textcolor{red}{3} of this book.

        Currently, there is no firm evidence against the stalling of all SMBHBs, although this may soon change if the recently detected evidence for a GW background can be definitively tied, at least in part, to a population of inspiralling SMBHBs (see \S\ref{SubS:GWs}). In addition, there are multiple, realistic astrophysical processes that could allow the binary to breach the final parsec:
            \begin{itemize}
            \item \textbf{Stars} in non-spherical distributions can torque the orbits of individual stars to lower-angular momentum orbits that interact with the binary. Triaxial stellar distributions have been shown to be capable of such ``loss-cone''\footnote{Referring to the shape of orbital energy and angular momentum space that results in strong interactions with the binary.} refilling at a high enough rate to merge the binary in a Hubble time \citep{2002MNRAS.331..935Y,2004ApJ...606..788M,2011ApJ...732...89K,2011ApJ...732L..26P}. 
            \item \textbf{Gas} can be torqued to the centers of galaxies following galactic merger \citep[\eg,][]{Barnes:1996, Barnes:2002} and could facilitate binary merger by removing energy and angular momentum from the binary orbit. 
            \item \textbf{Other} processes/mechanisms can also push the binary to small separations. Interaction of the binary with an incoming massive perturber \citep[\eg,][]{2007ApJ...656..709P} or star cluster \citep[\eg,][]{Bortolas_starclsuter+2018} may facilitate merger, or the interaction of the nuclear star clusters surrounding each SMBH may be important for deciding the fate of the SMBH pair \citep[\eg,][]{Ogiya+2020}. Even if all other processes fail, a subsequent galaxy merger could introduce a third black hole \citep[\eg,][]{Ryu:2018,Bonetti_TripIV+2019}.
            \end{itemize}
    So the final-parsec problem as originally posed is no longer a problem. However, the question of how SMBHBs do or do not come together, the \textit{supermassive merger problem} is still very much open. It is not a problem in devising methods for how to get SMBHBs to merge, but rather in understanding how it is done in nature: how do SMBHBs interact with their astrophysical environments and do, or do not come together.
    A solution to the \textit{supermassive merger problem} requires further modeling of SMBHB-environmental interactions, and, perhaps more importantly, it requires definitive evidence for or against SMBHBs in the intermediate, sub-parsec stages of evolution. How to find such evidence is the topic of this review.

\subsection{\large  Connection to Population Predictions}
\label{S:pop_models}
To establish the connection between the above physical processes describing binary formation and evolution and the expectations for a discoverable SMBHB population, we briefly present a mathematical model describing the population and its evolution.
We write a continuity equation describing the evolution in time (and thus in redshift) of the number of SMBHBs per orbital parameters. For demonstration, we write the SMBHB number density $\Dam$ per orbital frequency $f$ and total binary mass $M$, as a function of time $t$, i.e. $\Dam(f,M,t)$, valid from the formation of a bound binary down to merger,
\begin{eqnarray}
\partial_t \Dam + \partial_f\left(\dot{f} \Dam \right) 
+ \partial_M\left(\Mdot \Dam \right) = \Src, \qquad \Dam \equiv \frac{d^2n}{df dM} ,
\label{eq:SMBHB_cont}
\end{eqnarray}
where $\dot{f}$ and $\Mdot$ are rates-of-change of the binary orbital frequency and total binary mass, which generally depend on time and orbital parameters, and $\Src$ describes the rate at which new hard binaries are formed. One can solve this equation with the desired complexity in choices of initial redshift distribution, black hole growth, and orbital evolution. The result is a prediction for the population, $\Dam$.

We derive a relatively simple population model by linking the EM-bright SMBHB population to the quasar population following \citet{ SoyuerGWDop+2021} and relying on similar, previous approaches \citep{Sesana+2005, ChristianLoeb:2017}. We assume:
\begin{itemize}
    \item A constant fraction $\eta_{\bin}$ of all quasars are triggered by a galaxy merger that eventually results in the formation and merger of an SMBHB over the course of the quasar lifetime $t_Q$. 
    \item The quasar redshift and mass distribution traces the SMBHB merger and total binary mass distribution. Then the differential volumetric merger rate is 
   \begin{equation}
   \frac{ \mathrm{d} \mathcal{R}}{\mathrm{d}M} = \frac{\eta_{\bin}}{\tau_Q} \frac{\mathrm{d}^2N_Q}{\mathrm{d}L\mathrm{d}V} \frac{\mathrm{d}L}{\mathrm{d}M}
   \label{Eq:dRdMSMBHB}
    \end{equation}
    where $d^2N_Q/(dLdV)$ is an observationally determined quasar luminosity function, and we relate the quasar bolometric luminosity $L$ to the binary mass $M$ by choosing a distribution of accretion rates normalized in terms of the Eddington rate \citep[\eg,][]{Shankar_fEdds+2013}.
    \item The population is in steady state, consists only of circular orbits, and because we tie the binary mass to the quasar central mass, we do not include mass growth ($\Mdot=0$). Then at scales below where the source term contributes\footnote{The source function does not affect the solution for binaries at separations closer than a minimum separation scale at birth \citep[\eg,][]{ChristianLoeb:2017}. If however we can probe the population on such a scale, \eg, with observations of parsec-scale binaries and their progenitors such as dual AGN (\S\ref{S:Obs_Macro}), then we may also access this birth term.}, Eq. (\ref{eq:SMBHB_cont}) has solution $\dot{f}\Dam = \rm const. =  \mathrm{d} \mathcal{R} / \mathrm{d}M$.
\end{itemize}
With these simplifying assumptions, we can write the solution for the binary number density as
\begin{equation}
   n_{\bin} = \int{ \frac{ \mathrm{d} \mathcal{R}}{\mathrm{d}M} \frac{\mathrm{d}f}{ \dot{f}} \mathrm{d}M},
\end{equation}
where $\mathrm{d} \mathcal{R}/\mathrm{d}M$ is the volumetric merger rate per total binary mass for a specified population, and where we have further assumed that binary mass ratios are fixed over the observation time.

Then the total number of quasars within redshift $z$, total mass less than $M$, and orbital frequency above $f$ is
\begin{equation}
    \label{Eq:RateInt}
  N_{\bin}(>f) = 4 \pi \frac{\eta_{\bin}}{t_Q} \int\limits^z_0 \int\limits^{M}_{0} \int\limits^{f_{\rm{ISCO}}}_{f}{ \frac{\mathrm{d}V}{ \mathrm{d}z} \frac{\mathrm{d}^2N_Q}{\mathrm{d}L\mathrm{d}V} \frac{\mathrm{d}L}{\mathrm{d}M} \frac{f}{ \dot{f} } ~ \mathrm{d}\mathrm{log} f \mathrm{d}M \mathrm{d}z} ,
\end{equation}
where $\mathrm{d}V/\mathrm{d}z$ is the angle integrated cosmological volume element in a flat universe \citep{HoggCosmoDist:1999}. The integration must be limited to the range of applicability of the quasar luminosity function (\eg, a luminosity cut).

It is useful to divide by the total number of quasars in the same domain to estimate the fraction of AGN that harbour SMBHBs in the given mass, frequency and redshift range. This gives,
\begin{equation}
    \label{Eq:Ntres}
  \frac{N_{\bin}}{N_Q}(>f) \approx \eta_{\bin} \frac{\left< t_{\res} \right>}{t_Q},
\end{equation}
where $t_{\res} \equiv f/\dot{f}$ is the binary orbital parameter-dependent residence time (i.e. the time that the binary spends at a given orbital period), and the angle brackets denote a weighted average over the quasar mass and redshift distribution. This says that the longer the binary residence time, the higher the fraction of AGN that host SMBHBs with orbital frequency greater than $f$, and so the more abundant they should be in a survey of quasars.  Note that this argument holds even if AGN activity is intermittent \citep{GouldingAGNact+2018} over the binary lifetime, as long as the AGN has active stages that last for $\gtrsim t_{\res}$ when the binary is at orbital frequency $f$ \citep[see also][]{HKM09}; and the residence time $\sim10^5 \mathrm{yr} (f^{-1}/5.1\mathrm{yr})^{8/3} (M/10^8\Msun)^{-5/3}$ can be short compared to expected AGN lifetimes $\sim10^7-10^8$~yr \citep{PMartini:2004, GouldingAGNact+2018}. That the binary is bright during these times is of course an assumption of the model and subsumed into $\eta_{\bin}$.
For application to observations, one must also apply a flux limit which restricts the accessible luminosity and redshift space.

This approach has been used in a number of works to estimate properties of the SMBHB population \citep[\eg,][]{HKM09,  DOrazioLoeb_Gaia:2019, XinHaiman_LSSTshort:2021, Casey-Clyde+2022}. Further complexifications allow different orbital parameters to evolve, include other hardening mechanisms such as interactions with gas and stars \citep[\eg,][]{Mingarelli+2017, DOrazioLoebVLBI:2018, Bortolas:2021}, include non-trivial source functions, or rather than anchor to an observed quasar luminosity function, simulate the entire process with sub-grid prescriptions for SMBHB evolution painted onto the outputs of cosmological simulations \citep[\eg,][]{Kelley+2017a, Kelley+2017b, KelleyDop+2019, KelleyLens+2021}.

\section{\large  \large Observational Search Methods}
\label{S:IIIObs}

\begin{figure}
\begin{center}
\includegraphics[width=\textwidth]{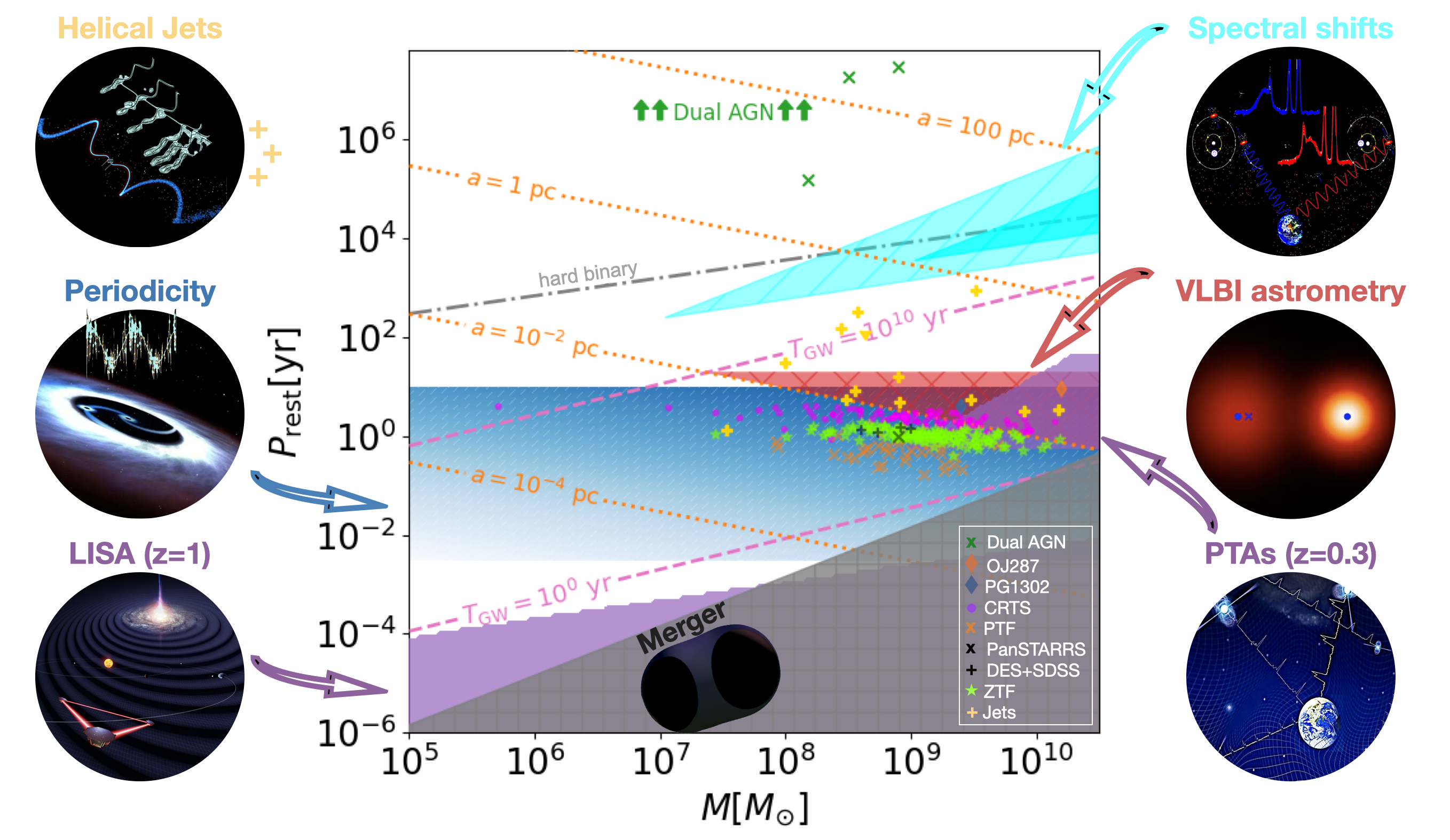}
\caption{
\textit{A visual guide to the Chapter:} Observational signatures of SMBHBs across population parameter space. The grey dot-dashed line is the hard-binary limit \citep[\S\ref{S:IIphys_procs};][]{Milosavljevic:2003:FPcP} -- below this line we consider SMBH pairs to be bound SMBHBs. Above this line the green x's represent selected dual AGN candidates from \citet{Koss+2023} and \citet{Goulding+2019}, as well as the most compact imaged AGN pair (7.3pc) from \citet[][]{Rodriguez+2006} (see \S\ref{SubS:Dual} and Figure~\ref{Fig:dual}). The cyan region denotes where kinematically offset broad lines are detectable following \citet{KelleyBLR:2021} and for binary mass ratio choices of $q=0.05$ (larger triangle) and $q=1$ (see \S\ref{SubS:BLRs}). The blue-white shaded region is where periodic lightcurve candidates are detectable with present and upcoming optical time-domain surveys (see \S\ref{S:PLC}). The darker shading indicates where the binary residence time is longest (larger $P$) and the accreting binary would be the brightest (larger $M$). Magenta dots \cite{Graham+2015b}, orange x's \citep{Charisi+2016}, the black x \citep{LiuGezari+2019}, black +'s \citep{Chen_DES_PLCs+2020}, and chartreuse stars \citep{Chen_ZTF_PLCs+2022} denote photometric variability candidates from CRTS, PTF, Pan-Starrs, DES, and ZTF respectively (see \S\ref{SubS:PLCsearches} and \S\ref{apendix:Phot}). The thick yellow \textbf{+}'s indicate the first 13 helical jet candidates listed in \S\ref{apendix:jets}, as identified by the methods described in \S\ref{S:JetMorph}. The red shaded triangle shows where direct orbital tracking techniques are effective (see \S\ref{S:DirImg}). The purple shaded regions denotes where LISA (bottom left) and PTAs (middle right) are sensitive at the indicated redshifts. Unless otherwise specified, the binary mass ratio is assumed to be $q=0.3$.
}
\label{Fig:BHBDemo}
\end{center}
\end{figure}

In the previous section we discussed how to use models of binary orbital evolution ($\dot{f}, \dot{M}, ...$) to predict populations ($\Dam \equiv \partial^2 n /\partial f \partial M$). To envision the role of observations, consider an opposite approach to Eq. (\ref{eq:SMBHB_cont}), where we sample $\Dam$ through observations of SMBHBs at different stages of their lives, and thus constrain $\dot{f}$ and $\Mdot$, allowing us to constrain the physics of SMBHB orbital decay and growth. We are not aware of a theoretical study determining the sufficient or optimal sampling of $\Dam$ through observations of SMBHB candidates to make non-trivial conclusions about $\dot{f}$, $\Mdot$, or other orbital change rates. However, multiple studies have applied candidate demographics towards gaining insight into underlying orbital decay mechanisms \citep[\eg,][]{PG1302MNRAS:2015a, Charisi+2016, Goulding+2019}, though with the caveat that the population does not consist of confirmed SMBHBs, or consists of a few (or even one) systems. Hence, we now review the existing evidence for, and various techniques to find, SMBHBs from the largest scales, down to the smallest.
Figure \ref{Fig:BHBDemo} serves as an overview of the methods and candidates that we review in this Section.

\subsection{\large  How and Where to Look: Population Expectations}
\label{S:popexp}
    
    The idea that SMBHBs exist in galactic nuclei and ideas for finding them have been around for over 40 years \citep{BBR:1980}. So why do we still have only circumstantial evidence for their existence at sub-parsec scales, and what would constitute definitive evidence?
    
    {\em Where to look?} To find electromagnetic evidence, an SMBHB must be electromagnetically bright. 
    In the majority of cases, this suggests that we should target a special type of galactic nuclei, namely active galactic nuclei (AGN)\footnote{Note that throughout the paper we use the terms AGN and quasar interchangeably -- quasars are a sub-class of AGN characterized by their high luminosity.}, in which one of both SMBH(s) is(are) accreting (see, however, \S ~\ref{SubS:CoredE} for binary tidal disruptions in \S\ref{SubS:HydroVar}, which do not require an AGN). So our search will be guided by the distribution of AGN and the fraction which harbor an SMBHB in a range of orbital separations for which a specific observational strategy is sensitive (see Figure~\ref{Fig:BHBDemo}). 
    An estimate for this fraction will also guide searches. In \S\ref{S:pop_models} we presented a model useful for estimating this in terms of binary residence time and the quasar lifetime (\ie, length of time for which the AGN is active). If we assume decay of circular orbits due to GW emission and make a point evaluation (simply for illustrative purposes here) to remove the average over the quasar luminosity function in Eq. (\ref{Eq:RateInt}), then we can estimate the fraction of quasars containing SMBHBs with orbital periods shorter than a fixed period $P\equiv f^{-1}$, as, 
    \begin{equation}
       \frac{ N_{\bin} }{N_Q} \approx 10^{-3} \eta_{\bin} \frac{(1+q)^2}{4q} \left( \frac{P}{5.1 \mathrm{yr}}\right)^{8/3} \left( \frac{M}{10^9 \Msun} \right)^{-5/3} \left( \frac{t_Q}{10^8 \mathrm{yr}}\right)^{-1},
    \end{equation}
    or equivalently
    \begin{equation}
       \frac{ N_{\bin} }{N_Q} \approx 10^{-3} \eta_{\bin} \frac{(1+q)^2}{4q}  \left( \frac{P}{5.1 \mathrm{yr}}\right) \left( \frac{N_a}{150} \right)^{5/2} \left( \frac{t_Q}{10^8 \mathrm{yr}}\right)^{-1},
    \end{equation}
    where $N_a\equiv a /(2GM/c^2)$ is the separation expressed in Schwarzschild radii, $q$, and $M$ are the binary mass ratio and total mass, respectively \citep[see also][this is the same as their Eq. (50) if $\eta_{\bin}=1$]{HKM09}. The choice of orbital periods of $\mathcal{O}$(yr) is appropriate for the baselines of modern time-domain surveys. For reference, a binary with total mass $10^9 \Msun$ and a $5.1$~yr orbital period has a separation of $0.014$~pc.

    Hence, to obtain a sample of $100$ SMBHBs, our simple model predicts that one must survey $\gtrsim 10^5/\eta_{\bin}$ quasars.  If the aim is to capture orbital variations (\eg, \S\ref{S:SpecSigs} or \S\ref{S:PLC} below), then any observing program that can survey the required number of quasars must also operate for at least one orbital period, but likely more in order to build the signal significance.
    Adopting the quasar luminosity function from \citet{Hopkins_QLF+2007}\footnote{The pure-luminosity-evolution, double-power-law with redshift dependent slopes. See the last row of Table 3 labeled ``Full.''}, we find that the number of AGN/quasars between bolometric luminosities of $10^{42}-10^{48.5}$erg/s does not rise above $10^5$ ($10^6$) until $z\sim0.05$ ($z\sim 0.1$). This suggests that a full sky survey of quasars must monitor beyond $z\sim 0.1$, for a decade ($\sim 2P$).

    The angular separation of the binary on the sky, $\theta_{\orb}$, and the prospect of resolving it, depends on the distance to the binary, and its semi-major axis. 
    The angular diameter distance at $z=0.1$ is $D_{\rm{A}}(z=0.1) = 380$~Mpc, while the majority of quasars will be at larger distances, with the quasar luminosity function peaking at $z=2$, near distances of $D_{\rm{A}}(z=2) = 1.7$~Gpc.
    Hence, in the best case scenario of a plausibly nearby, massive binary on a long period (but human-timescale) orbit, the angular scale of the orbit is,
    \begin{equation}
    \theta_{\orb} = \frac{a}{D_A(z)} \approx 7.8  \mu \mathrm{as}  \left( \frac{P}{5.1 \mathrm{yr}} \right)^{2/3}  \left( \frac{M}{10^9 \Msun} \right)^{1/3} \left(\frac{D_\mathrm{A}}{200 \mathrm{Mpc}} \right),
    \label{Eq:thetBin}
    \end{equation}
    which is a factor of $\sim2.5$ times smaller than the diffraction limited resolution of an Earth sized millimeter-wavelength very-long baseline interferometer (VLBI), \eg, the Event Horizon Telescope. Hence, while directly resolving orbits is not impossible (\S\ref{S:DirImg}), it is at the limit of present capabilities, and thus the vast majority of SMBHB search techniques rely on indirect detection methods.

   {\em What to look for?} Viable observational signatures for identifying accreting SMBHBs must be clearly discernible from the signatures of their single SMBH counterparts. While the observational characteristics of accreting singles are often associated with normal quasar behavior, one cannot yet be certain if ``normal quasar behavior'' is the same as normal accreting-single-SMBH behavior. However, the above arguments, which suggest that sub-parsec SMBHBs are rare, also suggest that the quasar population on average has properties that represent the normal mode of single SMBH accretion, and that, if unique identifiable signatures of SMBHBs exist, they will be outliers among the quasars. The caveats here are that (1) our handle on the SMBHB population is orders of magnitude uncertain (although the recently discovered evidence for a nanohertz GW background by pulsar timing arrays (PTAs) already provides useful constraints, which will improve as PTAs better characterize the GW background spectrum--see \S\ref{SubS:GWs}), so in an extreme case, what we know about normal quasar activity could very much be tied to SMBHBs, and (2) most SMBHBs may simply not state their presence in the universe, or act observably very similarly to single SMBHs for most of their lives.

   So how do we get beyond this? We must further develop {\bf (1)} our understanding of the signal for which we are searching while {\bf (2)} better characterizing the noise, in this case the behavior of ``normal quasars'' across temporal, wavelength, and spatial scales. For {\bf (2)} we stress the importance of understanding the noise in order to measure a signal, but leave that for a different review. For {\bf (1)}, the topic of this chapter, we aim to identify signatures of SMBHBs that are {\em (i)} ubiquitous, \ie, arising in a large fraction of SMBHB systems, and {\em (ii)} uniquely identifiable with SMBHBs. For the proposed signatures discussed below, the reader will see that it is difficult to fulfill both criteria, and when possible, it often requires observational techniques at the edge of our current abilities. 
   We now review these approaches, paying attention to ubiquity and uniqueness, highlighting candidate systems when applicable, and compiling a candidate list organized by search method in \S\ref{Sec:obs_summary}.\footnote{We do not attempt a complete list of present-day candidates, but note that the existence of a living database of candidates as they gain more evidence for or against their candidacy would be useful to the field (\textcolor{cyan}{\eg, Sydnor et al, in prep}).}

\subsection{\large  Macro-Scale Signatures and Evidence}
\label{S:Obs_Macro}

At scales of kiloparsecs down to $\sim10$\,parsecs, we do have direct evidence for accreting SMBH pairs (dual AGN, \S\ref{SubS:Dual}). On these larger scales we also have circumstantial evidence that stellar interactions drive the two SMBHs together, potentially causing the brightness deficits observed in the cores of galaxies which are likely the products of major galaxy mergers (cored ellipticals, \S\ref{SubS:CoredE}).

\begin{figure}
\begin{center}
\includegraphics[scale=0.7]{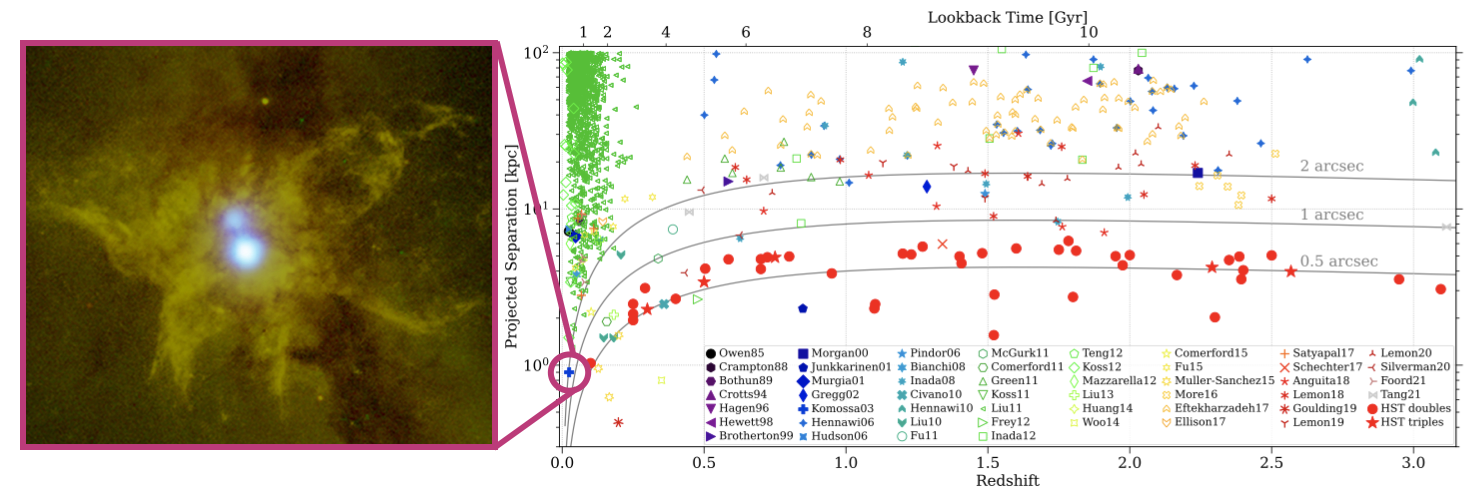}
\caption{{\em Left:} One of the first dual AGN detected in the galaxy NGC 6240 with blue showing the X-ray emission from the SMBHs overlaid on optical image. {\em Right:} Dual AGN candidates plotted in separation and redshift space. Adapted from \cite{Chen_Varstrometry+2022}.}
\label{Fig:dual}
\end{center}
\end{figure}

\subsubsection{\large  Dual AGN}
\label{SubS:Dual}
The first stages of galaxy mergers and the pairing process of their SMBHs have been repeatedly observed as dual AGN. These are galaxies that have two active SMBHs, separated by kiloparsecs (and in a few cases by hundreds or tens of parsecs) and a few dozen of those have been resolved across the electromagnetic spectrum. Figure \ref{Fig:dual} provides a representation of known dual AGN (and dual AGN candidates) adapted from \citet{Chen_Varstrometry+2022}.

Despite the observational success of the last decade, dual AGN overall remain relatively rare and challenging to detect. Candidates are typically selected either serendipitously or through systematic searches and then verification methods are employed to confirm or exclude the presence of two SMBHs in the candidate galaxy. These typically involve observationally demanding multi-wavelength follow-ups. One of the main challenges in the quest for dual AGN is the inherent observational trade-off between large field of view (required to build large samples of AGN) and high resolution (crucial for confirming the dual AGN). For instance, mid-infrared (IR) surveys, like WISE, have covered the entire sky detecting millions of AGN but their resolution is extremely limited, while at the other end, VLBI has exceptional resolution but cannot survey the sky and is only suitable for detailed follow-ups. Similarly high-resolution telescopes like Chandra and HST have only covered a small fraction of the sky. Another major challenge is the high degree of bias in the detected samples, which is hard to account for. Since systematic searches typically pre-select candidates by compiling a large sample of AGN, and then identifying a promising dual signature, they can be significantly biased (\eg, mid-IR methods preferentially select star-forming, dusty AGN, whereas X-ray methods prefer AGN with lower gas content, \citealt{2017ApJ...835...27A,2020ApJ...888...78S}). Hence, the observed sources are not necessarily representative of the underlying population, which complicates the connection with the galaxy mergers and the occurrence of binaries at smaller separations.

We emphasize, however, that the detection of dual AGN has significantly advanced our understanding of galaxy formation and evolution and can offer insights into observational strategies to detect sub-parsec binaries. For this reason, here we briefly summarize some of the observational highlights, but we refer the reader to \citet{DeRosa+2019} for a detailed review.

In one method, candidates are selected in the vast spectroscopic database of SDSS by searching for double-peaked emission lines (e.g., primarily [OIII], but also [NeV] and [NIII]), which reflect the motion of the AGN in the common gravitational potential, assuming that each AGN carries its own narrow emission line region \citep{Wang+2009, Liu_DualNarrowI+2010, Comerford+2013}. However, because this signature is not unique and other kinematic effects, such as outflows or rotating gas disks, can result in similar profiles \citep{KingPounds:2015,2016ApJ...832...67N}, additional data are required to confirm the existence of two SMBHs in the host galaxy. Detailed optical imaging with the Hubble Space Telescope and Keck, along with integral-field unit (IFU) spectroscopy, have confirmed that $\sim$2\% of the candidates are indeed dual AGN \citep{Fu_DualAGN+2011,McGurk+2015,Liu_DualAGN+2018}. We note that IFU spectroscopy can also provide a good alternative for candidate identification, but it is typically demanding and is only accessible from a limited set of telescopes. JWST is extremely promising in this regard (since MIRI and NIRCAM are equipped with IFU) and is expected to rapidly increase the number of known dual AGN.

Beyond the optical band, X-ray imaging and spectroscopy have also uncovered several dual AGN systems, especially in obscured galaxies. X-ray observations have either targeted candidate systems described above or have focused on fields with extensive coverage (e.g., COSMOS, Chandra Deep Fields) leveraging multi-wavelength data to uncover dual AGN candidates. Additionally, a number of dual AGN have been discovered serendipitously \citep{Komossa_NGC6240+2003, Koss+2011, Foord_DualAGNcand+2019}.
Currently, Chandra has the best spatial resolution of one arcsecond in X-rays and can resolve dual AGN systems with separations down to $\sim$8\,kpc at $z=1$ (or $\sim$1.8\,kpc at $z=0.1$). Note that Chandra's resolution decreases off axis but can be improved with advanced statistical methods like BAYMAX \citep{Foord_DualAGNcand+2019}.

The highest overall spatial resolution of order milli-arcseconds can be achieved in radio bands with VLBI, allowing the detection of dual AGN down to 10 parsecs and even smaller in the local universe (see also \S\ref{S:DirImg}). One limitation to this is that both AGN should be bright in radio wavelengths, which may be rare; in general radio-loud AGN are <10\% of the population (\eg, see \citealt{2015ApJ...813..103M} on dual AGN verification with radio follow-ups).
In addition, because VLBI has a narrow field of view, it is most suitable to study pre-selected promising individual objects or small samples of candidates. This method has returned the record-holding binary at a projected separation of 7.3pc \citep{Rodriguez+2006}, in which the relative motion of the two SMBHs has also been observed \citep{2017ApJ...843...14B}.

 Detection of sub-kiloparsec dual AGN (10pc-1kpc) can also be achieved through precision astrometry paired with variability \citep[\eg,][]{Popovic_GaiaVarstromtry+2012}. This method, recently referred to as varstrometry, tracks the photocenter of light coming from both AGN and host-galaxy light. As one or both AGN vary in brightness due to intrinsic (or possible binary induced) variability, the photocenter changes, allowing one to probe photocenter offsets between components \citep{Shen_Varstrometry+2019, Hwang_Varstrometry+2020, Chen_Varstrometry+2022}.

\subsubsection{\large  Cored Ellipticals}
\label{SubS:CoredE}
As discussed in \S\ref{S:IIphys_procs}, an important mechanism for hardening SMBHBs is stellar interactions, whereby SMBHB angular momentum and energy is traded to stars on initially centrophilic orbits, kicking them out of the central region of the stellar core which surrounds the SMBHB. For this reason, \citet{BBR:1980} first pointed out that the stellar core of a galaxy that recently harboured an SMBHB merger should have a deficit of stars within the hardening radius of the binary \citep[grey-dot-dashed line in Figure \ref{Fig:BHBDemo} where the binary orbital velocity matches the stellar velocity dispersion, see also,][]{MM_GalCores+2002, Ebisuzaki_CoreEll+1991}. Such stellar cores are indeed found in the brightness profiles of some elliptical galaxies \citep{Lauer_CoreEllObs:1985, KormendyHo2013}, though it is difficult to provide direct proof that cored ellipticals are created by SMBHB scouring. For example, AGN feedback from single-SMBH accretion has been proposed \citep[\eg,][]{Martizzi+2013} as an alternative culprit. However, convincing evidence for the binary picture has been mounted. Namely: 
{\bf (1)} Mass deficits predicted by binary+stellar hardening theory match the measured central SMBH mass in cored ellipticals \citep{Merritt:2006, KormendyBender:2009}.
{\bf (2)} The core size matches the SMBH sphere of influence, derived from measured SMBH and stellar properties \citep{Thomas+2016}.
{\bf (3)} Anisotropic velocity dispersions observed in cored ellipticals could be caused by preferential removal of stars on highly radial orbits during the scouring process \citep{Rantala+2018}.

 \begin{center}
 \begin{tcolorbox}[width=0.98\textwidth,colback={white},title={{\bf Pros and Cons: Macro-Scale Signatures}}, colbacktitle=gray,coltitle=black] 
 \textbf{Pros:}
 \begin{itemize}
    \item We have successfully observed dozens of double AGN sources, and cored ellipticals.
     \vspace{-5pt}
     \item Prospects with the James Webb Space Telescope or future high energy telescopes with high resolution like LynX are promising.
 \end{itemize}
 \textbf{Cons:}
 \vspace{-5pt}
 \begin{itemize}
     \item Multi-wavelength follow-ups for dual AGN are observationally demanding.
     \vspace{-5pt}
     \item Dual AGN samples are not complete and selection effects hinder the extraction of physical quantities like the rate of dual AGN in galaxies or relevant timescales.
     \vspace{-5pt}
     \item Due to uniqueness problems, it is not clear if one can tie the rate of stellar hardened SMBHB pairings to the cored ellipticals.
 \end{itemize}
 \end{tcolorbox} 
 \end{center}

\subsection{\large  \large Spectral Signatures}
\label{S:SpecSigs}

        The idea for using spectral features to detect SMBHBs was prompted first by observations of broadened emission lines offset in redshift from narrow emission lines in the then newly discovered quasars \citep[][very likely inspired by a history of using spectra to identify binary stars in astronomy]{Komberg:1968, Gaskell:1983}. This has lead to a long tradition of searching for offset, variable, and double peaked broad emission lines as evidence for SMBHB orbital motion, which we discuss in \S\ref{SubS:BLRs}. In addition to broad line signatures, advances in modeling accretion onto binaries has lead to other proposed spectral signatures arising from more indirect outcomes of a binary orbital motion, such as clearing of gaps and cavities in circumbinary accretion disks. We discuss signatures of these processes occurring much closer to the SMBHs in subsection \S\ref{SubS:CBDspectra}.

        \subsubsection{\large  Broad Line Features and Variability}
        \label{SubS:BLRs}

\begin{wrapfigure}{r}{0.5\textwidth} 
\begin{center}
\vspace{-20pt}
\hspace{-7pt}
\includegraphics[scale=0.5]{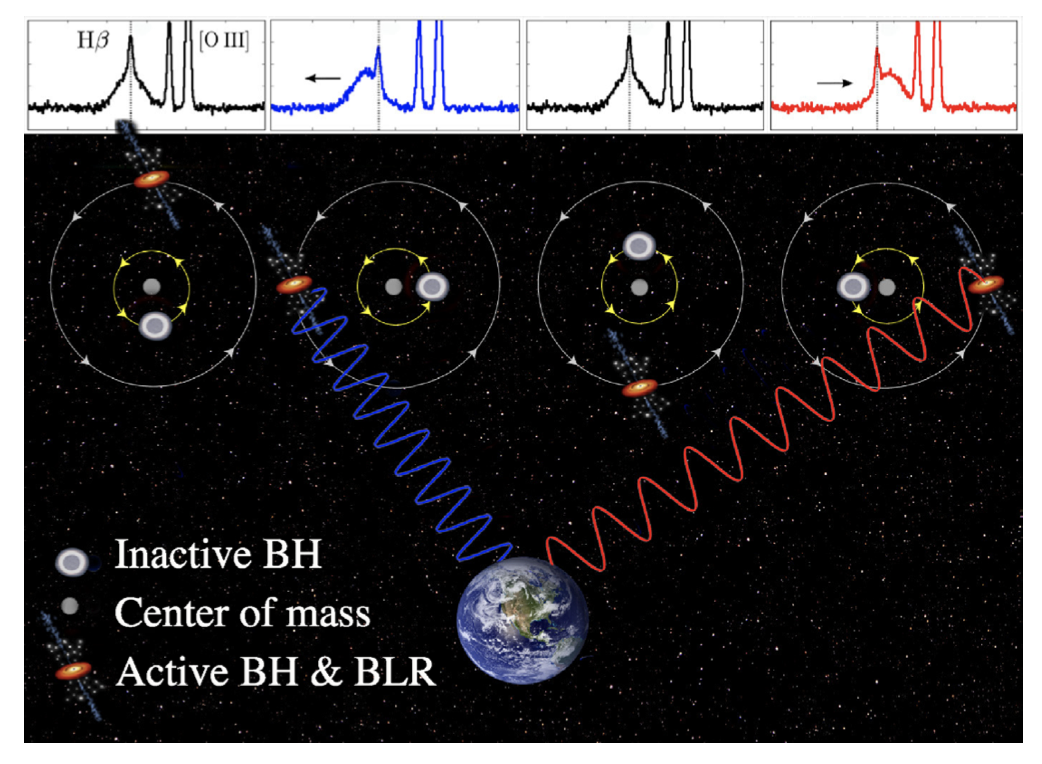} 
\caption{Cartoon showing the expected Doppler shift in broad emission lines, induced by the motion of the binary, assuming that only one SMBH is active and carries its own broad line region. The expected coherent modulation in radial velocity provides a required test for binarity but the timescales for this are long. From \cite{GuoShen_BLsearchIII+2019}.
}
\label{Fig:Spectroscopic}
 \vspace{-20pt}
\end{center}
\end{wrapfigure}

        Gas ionized by the central power source of AGN, i.e. the accreting SMBH(s), emits a multitude of emission lines at different atomic transitions. The width of the so-called broad emission lines is due to Doppler broadening caused by orbital motion of gas around the central mass, while emission from much more slowly orbiting gas farther from the central source is not greatly broadened, resulting in reference narrow lines \citep[see, \eg,][for a classic review of emission lines in AGN]{OsterbrockMatthews:1986}. Offsets of $\sim10^{3}$km/s (up to 5,000km/s) in redshift between broad and narrow line centers could suggest that the broad line region is coherently moving, because it is bound to a component of a binary and so red- or blue-shifted by the SMBHB orbital velocity (see Figure~\ref{Fig:Spectroscopic}). 

        Early studies (before 2010) focused on interesting individual sources with peculiar spectral features \citep{Komberg:1968,Gaskell:1983,StockFarn:_OX169:1991,Gaskell+1996,Bogdanovic2009candidate,Dotti+2009,BorosonLauer:2009,Decarli2010,Barrows2011,Bon+2012}. 
        Subsequently, the vast spectroscopic dataset of SDSS enabled many systematic searches for binary broad line signatures \citep{Tsalmantza+2011,Erac+2012,Shen_BLsearchI+2013,Ju+2013,LiuShen_BLsearchII+2014}, see also \S\ref{apendix:Spec}. These signatures can be roughly sorted into three categories:

        \begin{itemize}
            \item {\em Broad lines kinematically offset with respect to the narrow lines identified through single-epoch spectroscopy}. The narrow emission lines are produced at large distances from the galactic nucleus and are not affected by the presence of a binary, thus tracing the rest-frame of the galaxy. If one of the binary components accretes at a higher rate than the other and carries with it the only broad line region of the system, the broad emission lines will be displaced compared to the narrow lines, due to the orbital velocity of the SMBH \citep{Dotti+2009, Tsalmantza+2011, Erac+2012, LiuShen_BLsearchII+2014}.
            \item {\em  Broad lines varying in offset (or shape) over time identified through multi-epoch spectroscopy}. The broad line variability may arise in a binary system if each SMBH has its own broad line region or if only one of the SMBHs is accreting \citep{Shen_BLsearchI+2013, Ju+2013,Wang+2017}. This method has the advantage of detecting candidates at a variety of orbital phases, even when the line-of-sight radial velocity is zero (which would have been missed in single-epoch searches) since at that phase the line-of-sight acceleration is maximum. These searches attempted to provide the first constraints on the fraction of quasars hosting binaries, but the range is wide (from 1\% to almost all quasars being associated with binaries) and dependent on assumptions about the unknown physics of broad line regions.
            Additionally, most of the candidates identified from single-epoch spectroscopy were subsequently monitored to search for coherent changes in the radial velocities, as expected in the binary scenario \citep{Erac+2012,Decarli+2013, LiuShen_BLsearchII+2014,Runnoe+2015,Runnoe+2017}. Some of the candidates have evolving radial velocities consistent with the orbital motion of a binary or have led to constraints on the binary parameters \citep{Runnoe+2017,GuoShen_BLsearchIII+2019}, while others have already been ruled out due to lack of variability or due to radial velocity curves inconsistent with the orbital motion of a binary \citep{Wang+2017,GuoShen_BLsearchIII+2019}.
            \item {\em Double-peaked broad lines} may be caused by two broad line regions, each bound to one of the black holes in the binary. Multiple candidates have been considered over the years, especially in the earlier days of the field, \citep{StockFarn:_OX169:1991, EracHalp:1994, BorosonLauer:2009, Tsai+2013}, but follow-up studies disfavored the binary interpretation of these systems, \citep[\eg, see][for reviews]{Erac_Dblpeak_AGNdsk:2009, DeRosa+2019}, and they may be more simply generated by single AGN disks \citep[\eg,][]{Erac_CandRejec:1997, EracHalp_DblPeak:2003,LiuErac+2016, DoanErac+2020}. This signature is also challenging from a theoretical perspective; the proper balance of line width and offset required for such a splitting to be observable places strict requirements on the allowed orbital parameters \citep{ShenLoeb:2010,KelleyBLR:2021}.
        \end{itemize}

        Kinematically offset broad lines are detectable for binary separations ranging from $\sim 0.1$pc to $\sim 10$pc (see cyan region in Figure~\ref{Fig:BHBDemo}), while systems for which time changing offsets are detectable make up a smaller portion of parameter space.
        The binary separation must be small enough (and the orbital velocity large enough) to detect offsets due to orbital motion \citep{Erac+2012, Ju+2013, Pflueger+2018, KelleyBLR:2021}, and the binary must also be close enough to be gravitationally bound if Keplerian evolution of the line offsets is to be expected. The above signatures also require that the binary is wide enough that at least one of its components can hold onto a bound, luminous broad line region \citep{Runnoe+2015, KelleyBLR:2021}. From these considerations, \citet{KelleyBLR:2021} estimates that for contemporary observational capabilities, kinematic offsets are detectable for $\sim0.5\%$ of a mock SMBHB population, for those with total masses $\gtrsim 10^7 \Msun$ and separations between $0.1$ and $10$~pc. The cyan region in Figure~\ref{Fig:BHBDemo} depicts the portion of parameter space where these criteria are met (Eqs. (6) and (10) of \citealt{KelleyBLR:2021} with $f_{\Edd}=0.1$), assuming $10^3$km/s broad line offsets due to orbital motion of the secondary SMBH, and assuming mass ratios of $q=0.05$ and $q=1.0$ (large and small cyan triangles, respectively, in Figure \ref{Fig:BHBDemo}) to bracket the range of binary masses and periods that allow such offsets. Changes in the kinematic offsets will only be detectable for the shorter orbital period systems in the cyan region of Figure~\ref{Fig:BHBDemo}, estimated to constitute $0.03\%$ of the mock SMBHB population in \citet{KelleyBLR:2021}.
    
        Recent modeling has considered circumbinary broad-line regions which could exhibit time-dependent line profiles and centroids due to the changing illumination pattern of the central binary \citep{Bogdanovic+2008, ShenLoeb:2010, PG1302MNRAS:2015a, NguyenBogI:2016, NguyenBogII:2019, NguyenBogIII:2020, Ji_CBD_BLR+2021}. Such signatures would exist in a extended part of parameter space, at smaller binary separations, where an intact broad-line region need not be bound to either binary component. However, comparison of recent model line profiles by \citet{NguyenBogII:2019} and \citet{NguyenBogIII:2020} between a sample of SMBHB candidates and a control sample shows that the inferred putative binary parameters between the two sets are indistinguishable. Hence, such a diagnostic may be useful for inferring parameters of known SMBHBs, but not for identifying them.
        While predictions for the circumbinary-broad-line response are less certain and possibly not unique to an SMBHB, correlation with photometric variability could help with uniqueness arguments. Correlation of the changing circumbinary-broad-line shapes and magnitudes with the photometric variability of continuum emission (\eg, \S\ref{S:PLC}) could offer a promising handle for testing the binary hypothesis and mapping broad-line regions \citep{JiLuGe:2021}.

        \subsubsection{\large  Circumbinary Disk Spectral Features}
        \label{SubS:CBDspectra}
        Another class of possible spectral signatures from accreting SMBHBs arises from the inner binary accretion flow itself. The idea is as follows: the spectrum of a standard steady-state accretion disk around a single black hole is found from solving for the disk radial temperature profile and summing the multi-component black-body emission from each radial ring \citep{SS73}. In the case of circumbinary accretion, however, the temperature profile (and even the optically thick, blackbody assumption) can be modified. For example, the disk surface density snapshots in Figure~\ref{Fig:2DDens}, for binaries of different mass ratios, show that the binary can clear a low density ring or cavity in the disk. At first approximation, this could result in lack of emission at these radii, and possibly a deficit (or `notch') in the equivalent single-SMBH-disk spectrum \citep{GultekinMiller_SEDGaps:2012, Roedig_SEDsigs+2014}. Other consequences of the cavity include softer than expected spectral energy distributions, weaker broad lines \citep{Tanaka+2012}, or spectral edges caused by Lyman-$\alpha$ absorption at the inner edge of the cavity \citep{GenerozovZH:2014}.

        Following the above predictions, \citet{Yan_Mrk231_notch+2015} proposed an SMBHB candidate with separation of $\sim 3$~milli-parsec based on an observed optical-UV deficit in the spectrum of the nearby quasar Mrk 231 \citep[but see][who suggested that the observed spectral features are due to dust-reddening]{Leighly+Mrk231+2016}. \citet{Guo_CandSEDs+2020} searched for predicted spectral energy distribution truncations or peculiarities in a set of 138 candidate SMBHBs identified by periodic brightness variations (see \S\ref{S:PLC}). Six systems showed abnormally red, blue-truncated spectra that are consistent with empty-cavity circumbinary disk spectral energy distribution models. However, this fraction is consistent with the fraction of similar outliers in a control sample, and could also be explained via dust-reddening. \citet{Foord_CandCBDspec+2022} examined spectra for the SMBHB candidate SDSS J025214.67-002813.7, identified by periodic brightness variations \citep{Liao_MdotPLC_cand+2021}. Compared to standard AGN SEDs, this system exhibits excess soft X-ray emission in the $0.5-10$~Kev range and a blue deficit above 1400\,\AA. However, to explain the latter, a dust reddening model is preferred over a binary hypothesis.  
        
        In addition to reprocessing or adjusting single-disk emission, a more sophisticated treatment of the binary accretion problem will also result in an underlying spectrum that deviates from the single SMBH case. For example, binary torques will contribute a heating term to the energy balance that can change the disk temperature profile. Manifestations of this are evident in, \eg, \citet{Roedig_SEDsigs+2014}, which predicted an X-ray excess due to Compton cooling in shocks generated as streams from the circumbinary disk hit the disks around each SMBH (often referred to as circumsingle or mini-disks; see Figure~\ref{Fig:2DDens}). The numerical calculations of \citet{Farris:2015:Cool, Tang_LateInspCool+2018, WS_eccCool+2022, RyanMacFadyen:2017} lend further support for high-energy excesses due to gas dynamics and shocks in the inner region of the circumbinary disk and mini-disks, albeit due to different physical reasoning. 
        
        The latter works carried out viscous hydrodynamical simulations with simple optically thick radiative cooling. In reality, the optically thick assumption (and hence radiatively efficient blackbody cooling assumption) may break down in the cavity, and result in different cooling mechanisms and different spectra.  For example, \citet{PG1302Nature:2015b} argue that for the SMBHB candidate PG 1302-102 (see \S\ref{S:DopLens}), the accretion flow onto the putative binary should be a combination of radiatively efficient \citep[optically thick black body, \eg,][]{SS73} and advection dominated accretion flows \citep[\eg,][]{NarayanYi:1995}, allowing the possibility for more complex spectral signatures.
        While such composite binary accretion flows have not yet been simulated, \citet{dAscoli+2018, Gutierrez+2022} carry out general relativistic magneto-hydrodynamical (GRMHD) simulations near merger that implement an ad-hoc cooling mechanism which cools towards a target entropy and reconstructs spectral energy distributions assuming black-body emission in the bulk, optically thick regions of the disk and an inverse Compton scattering emissivity in the optically thin cavity. In these simulations the inverse Compton emission from the cavity and SMBH mini-disks results in a high-energy coronal type component of the SMBHB spectrum. \citet{Saade_Chandra:2020} provide a more detailed overview of predicted high-energy (X-ray) features of accreting SMBHBs, motivating searches for peculiar X-ray spectral indices or abnormal X-ray-to-optical flux ratios (compared to the known AGN population) as indicators of SMBHB accretion. To date, a handful of SMBHB candidates with X-ray and optical observations have been examined under this light, but no significant abnormalities have been found \citep{Saade_Chandra:2020, SpikeyHu+2020, Saade_NuStar+2023}.

        Finally, \citet{McKernan_FeKa+2013} show that it is not only the spectral energy distribution that can be affected by gaps and cavities in circumbinary accretion disks, but also the shape of the Fe K$-\alpha$ spectral line, which appears in X-rays (at 6.4-7keV, rest frame) and probes the inner regions of SMBH(B) accretion disks \citep{Fabian_FeKa+2000}. \citet{McKernan_FeKa+2013} predict oscillations in the line shape due to Doppler boosting as well as missing wings (red and blue extremes) of the line due to cavity or gap formation. Such features may be observable with next-generation X-ray observatories such as Athena \citep{ATHENA:2013}. No Fe K$-\alpha$ lines were detected in the Chandra follow-up observations reported in \citet{Saade_Chandra:2020}, which is typical for such AGN at the reported count rates.

 \begin{center}
 \begin{tcolorbox}[width=0.98\textwidth,colback={white},title={{\bf Pros and Cons: Spectral Signatures}}, colbacktitle=gray,coltitle=black]    
 \textbf{Pros:}
 \begin{itemize}
     \item Spectroscopic databases like SDSS have provided large samples of quasars allowing for systematic searches, which have revealed many promising candidates.
     \vspace{-5pt}
     \item The time-domain component of SDSS-V, Black Hole Mapper, is well suited to uncover more candidates.
     \vspace{-5pt}
     \item The most widely separated binaries detected with this method can be resolved with VLBI if both components are radio loud (See \S's \ref{SubS:Dual} and \S\ref{S:DirImg}).
     \vspace{-5pt}
     \item Long term monitoring can provide stringent constraints on the binary parameters and potentially exclude the binary hypothesis.
 \end{itemize}

 \textbf{Cons:}
 \vspace{-5pt}
 \begin{itemize}
     \item The signatures are not unique to SMBHBs.
     \vspace{-5pt}
     \item The more unique, Fe K-$\alpha$ signatures are much harder to observe for any but the brightest AGN.
     \vspace{-5pt}
     \item For the traditional (optical) broad line signatures in the beginning of \S\ref{SubS:BLRs}, binary periods are long (decades to centuries) and thus observing a full orbit is hard and confirmation of binarity challenging.
     \vspace{-5pt}
     \item This signature likely probes a small part of SMBHB orbital parameter space (Figure \ref{Fig:BHBDemo}).
 \end{itemize}
 \end{tcolorbox} 
 \end{center}

\subsection{\large  Photometric Signatures (Periodic Variability)}
\label{S:PLC}

The primary idea behind the periodic variability method is that the difference between accreting single SMBHs and binaries can be borne out from periodic brightness variations in the lightcurves of quasars powered by binary accretion \citep[][]{HKM09}. This periodicity is imprinted, in some way or another, by the periodic orbital motion of the gravitational two-body problem.

Regardless of the process which imprints periodicity, the timescale is set by the binary orbital period and its multiples. SMBHBs with masses between $10^7-10^{10}\Msun$, with $0.01$~pc separations, at redshift $z$, have observer-frame orbital periods ranging from $\sim 30(1+z)$ to $\sim1(1+z)$~yr, and can be observed out to high redshifts when accreting near the Eddington rate.\footnote{LSST of the Rubin Observatory should detect SMBHs with masses $>10^6\Msun$ accreting at $0.1$ Eddington for $z\leq 1$.} In this Section we review how periodic orbital motion can be imprinted on observed emission from binary accretion, the status of observational searches, and challenges in detecting such signals.

\subsubsection{\large  Hydrodynamical Variability}
\label{SubS:HydroVar}

Theoretical and numerical models of circumbinary accretion have shown that the time-averaged accretion rate onto the components of a binary matches that expected onto a single point mass of equivalent mass \citep[\eg,][]{DHM:2013:MNRAS, Farris+2014}. In cases studied so far, torques from the binary on the inflowing circumbinary gas do not inhibit accretion onto the individual components (see, however, \citealt{Ragusa_Mdot+2016}, but also \citealt{DittmannRyan_Hnu:2022}). Therefore, binary-powered accretion can be as bright as accretion powered by a single SMBH. 

\begin{figure}[ht]
\begin{center}$
\begin{array}{c}
\hspace{-10pt}
\includegraphics[scale=0.44]{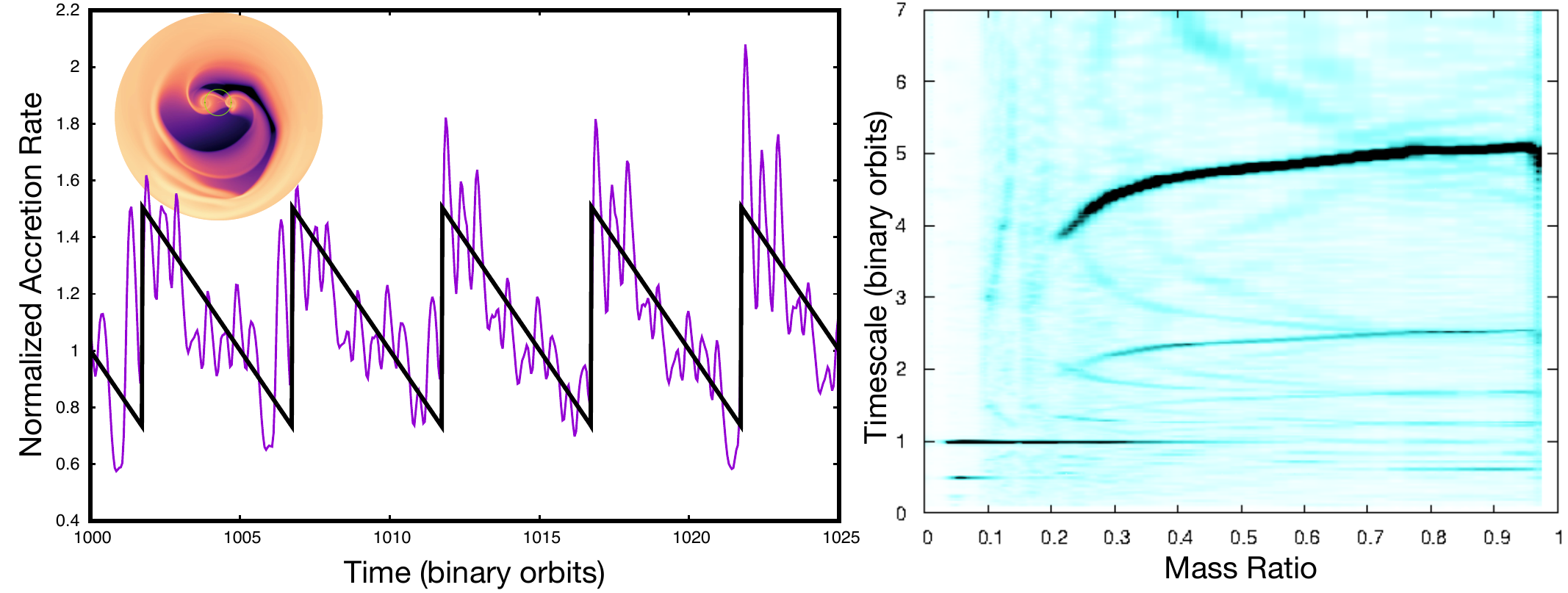} 
\end{array}$
\caption{
{\em Left:} Simulation measured accretion rate (purple) over 25 orbits for an equal mass binary on a circular orbit (adapted from \citealt{Duffell+2020}). The inset is a snapshot of log surface density in a disk simulation from which the accretion rate is measured. {\em Right:} A 2D periodogram of the accretion rate time-series for binary mass ratios spanning $0.01$ to $1.0$ (adapted from \citealt{Duffell+2020}). For low mass ratios, periodic variability at the binary orbital period is dominant. For near-equal mass ratios, periodicity at a combination of the binary orbital timescale, a five times longer periodicity, and harmonics all are present.
}
\label{Fig:Mdotvar}
\end{center}
\end{figure}

\begin{figure}[ht]
\begin{center}$
\begin{array}{c}
\hspace{-10pt}
\includegraphics[scale=0.44]{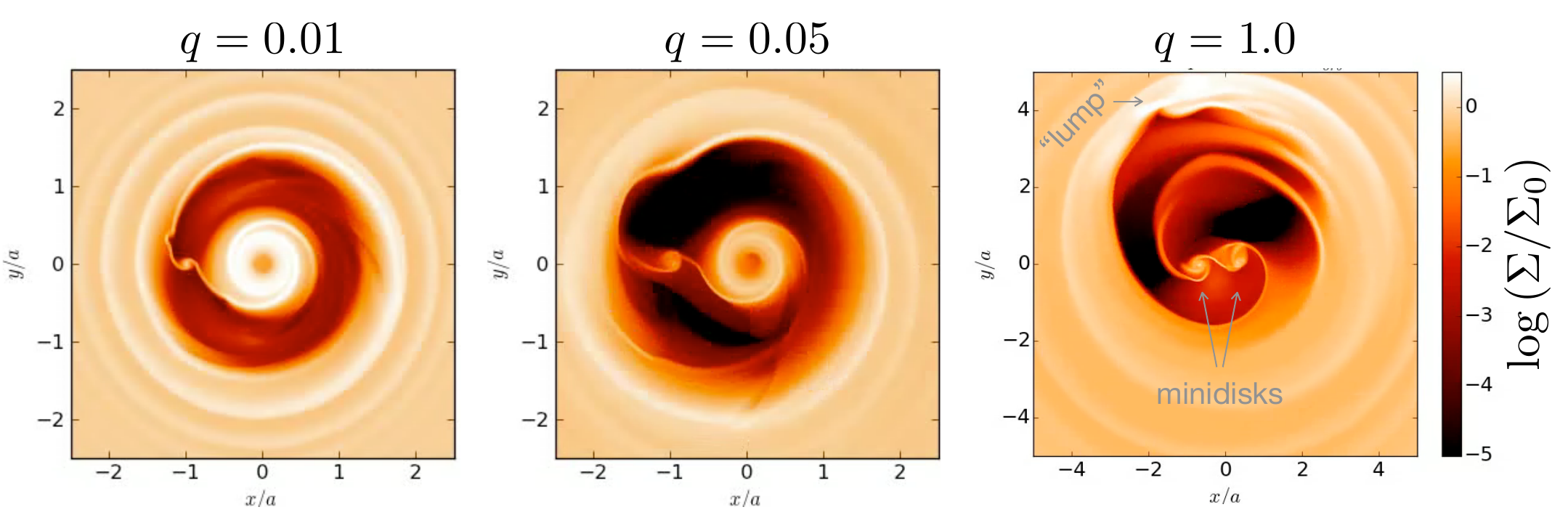} 
\end{array}$
\caption{
Log surface density snapshots of numerical calculations described in \citet{D'Orazio:CBDTrans:2016}, for different binary mass ratios $q=M_2/M_1\leq1$. These snapshots correspond to the three accretion-rate variability regimes for circular orbit binaries. The binaries are orbiting counter-clockwise.
}
\label{Fig:2DDens}
\end{center}
\end{figure}

Moreover, numerical studies show that the accretion rate onto the binary can be strongly modulated at the binary orbital period and multiples thereof. A landscape of accretion-rate variability arises with strong dependence on binary parameters. The left panel of Figure~\ref{Fig:Mdotvar} shows an accretion-rate time series for an equal-mass binary in a circular orbit, while the right panel shows a periodogram of accretion rate variability for a range of circular orbit mass ratios \citep{Duffell+2020}. 
For binaries on circular orbits, a trend with binary mass ratio $q$  
arises \citep{DHM:2013:MNRAS, D'Orazio:CBDTrans:2016}, and can be split into three regimes:
\begin{itemize}
    \item {\bf Three-timescale regime ($\mathbf{q \gtrsim 0.2}$):} The binary modulates the accretion flow at the orbital period and twice the orbital period by periodically pulling in streams from the circumbinary disk. For exactly equal mass binaries periodicity at the orbital period disappears due to symmetry. The majority of material from these streams is flung back out into the circumbinary disk maintaining a low-density central cavity \citep[\eg,][]{Tiede+2022}. This cavity becomes eccentric with a side near the binary, from which it pulls streams and accretes most strongly, and a side far from the binary into which the majority of stream material is ejected by the rapidly rotating binary potential. An overdensity (often referred to as the ``lump'') develops due to coherent impacts from streams that enter the cavity and are subsequently expelled back out hitting the far side of the cavity wall \citep[($2-4a$), see description in][]{DHM:2013:MNRAS}. This overdensity, labeled in Figure~\ref{Fig:2DDens}, builds up and then orbits at the local Keplerian frequency. Streams heading toward collision with the cavity wall, which re-enforce the overdensity, are also visible in the right panel snapshot in Figure~\ref{Fig:2DDens}. Once every $\approx 5$ binary orbits, the orbiting overdensity reaches the side of the cavity that comes nearest to the binary and causes an overfeeding, temporarily increasing the accretion rate onto the binary. As can be seen in the periodogram of Figure~\ref{Fig:Mdotvar}, this super-orbital periodicity dominates for near-unity mass ratios. This feature does not always arise (see below), but is seen in numerical calculations that use a range of techniques and include a range of physics, \eg, non-trivial thermodynamics, magneto-hydrodynamics (MHD), self-gravity, 2D and 3D \citep{MM08, DHM:2013:MNRAS, Noble+2012, ShiKrolik:2012:ApJ,  ShiKrolik:2015, D'Orazio:CBDTrans:2016, Munoz+2019, Moody+2019, RagusaCavs+2020, Noble+2021}.  
    \item {\bf Orbital-timescale regime ($\mathbf{0.05 \lesssim q \lesssim 0.2}$):} Below approximately $q=0.2$, the orbiting overdensity cannot be maintained and the super-orbital $\sim5P$ periodicity vanishes leaving only power at the orbital period and twice the orbital period (middle panel of Figure~\ref{Fig:2DDens} and right panel of Figure~\ref{Fig:Mdotvar}). Even though the lump formation seems to be robust across many different calculations, the criteria for its formation/destruction is an active area of research \citep[\eg,][]{Noble+2021, Mignon-Risse_lump:2023} and is likely dependent not only on the mass ratio, but on disk properties and thermodynamics \citep{Sudarshan_Penzlin_Cool+2022, WangCoolA+2022, WangCoolB+2022}.
    \item {\bf Steady regime ($\mathbf{q \lesssim 0.05}$):} For small mass ratios, the accretion-rate variability becomes low-amplitude and stochastic (left panel of Figure~\ref{Fig:2DDens}). This transition is coincident with a range of mass ratios near where the loss of stable co-orbital orbits occurs in the restricted three body problem ($q\geq0.04$), but also has dependence on disk properties \citep{D'Orazio:CBDTrans:2016, DittmannRyan+2023}. 
\end{itemize}

Accretion variability for eccentric, near equal mass ratio binaries is qualitatively similar to the circular orbit case for $e\lesssim 0.1$. For more eccentric binaries, however, the orbiting overdensity that generates the dominant super-orbital periodicity disappears, leaving an accretion rate dominated solely by the orbital timescale and its harmonics. The shape of the accretion rate modulation also changes with higher eccentricity, becoming more bursty, with a peak just preceding the time of binary pericenter \citep[][\textcolor{cyan}{D'Orazio \& Duffell, In prep.}]{Zrake+2021, MunozLai_PulsedEccAcc:2016, Dunhill+2015}. A systematic exploration of accretion rates for eccentric binaries with disparate mass ratios has been recently carried out by \citet{Siwek+2023}.

Much of the above picture has been realized using numerical viscous-hydrodynamical and MHD calculations with relatively simple or isothermal gas-cooling prescriptions and for a relatively narrow range of disk parameters. For example, the equilibrium disk aspect ratio which controls pressure forces, rarely deviates from the range $0.1-0.03$ in these calculations \citep{Tiede+2020}, despite expectations from radiatively efficient, steady-state disk theory that this value should fall below $0.01$ for accretion onto SMBHs \citep{SS73, SirkoGoodman:2003, ThompsonAGNDsk:2005}. However, recent parameter studies have found that the three-timescale regime above is relatively robust over a range of disk aspect ratios and viscous magnitudes \citep{DittmannRyan_Hnu:2022}, with \citet{DittmannRyan+2023} finding accretion rate variability down to $q=10^{-2}$ for thinner disks. These choices for disk properties control when, for example, the orbiting overdensity can be sustained and so control the boundaries between the above regimes. Even so, comparison with observed accretion rates onto stellar binaries has been carried out with reasonable success \citep{Tofflemire_MdotObs+2017}. Though extrapolation to the SMBH case may not be straightforward because in the stellar case the disk aspect ratio is expected to be closer to the widely simulated values of $\sim0.1$, and buffering of accretion rate variations by the mini-disks may be less important due to the larger physical extent of the stars or their magnetospheres. Finally, note that misalignment of the circumbinary disk with the binary will likely also alter this landscape of accretion variability \citep{Moody+2019, DouganNixon+2015, Smallwood+2022}.

While the accretion rate sets the spectrum and luminosity of a steady-state disk, it is not clear that the observed luminosity will exactly track variations in simulation-measured accretion rates. The accretion rate varies on the orbital timescale while a separate timescale controls how quickly this translates into a change in observed emission. This latter timescale may be set by the viscous time in the mini-disks surrounding each SMBH, the photon diffusion time required for a changing disk temperature to propagate to the disk photosphere, or other timescales relevant to the dominant form of radiative cooling. If these timescales are longer than the orbital period, then a resulting buffering of the accretion rate variability could result in diminished or no variability in the observed flux \citep[\eg,][]{BrightPasch_minidisks:2023}.

A few studies have solved the energy equation with a cooling prescription in order to track not just the accretion rate, but a (sometimes) frequency dependent luminosity. 3D GRMHD simulations of $\mathcal{O}(10)$ binary orbits near merger \citep{dAscoli+2018, Gutierrez+2022}, with the HARM3d code \citep{Noble+2009} include an ad-hoc cooling function and post-process the resulting disk radiation in the binary black hole spacetime, finding variability signatures in the disk emission. Other studies implement a simple optically thick black-body cooling function and find that the luminosity generally tracks the accretion rate \citep{Farris:2015:Cool, Tang_LateInspCool+2018, WS_eccCool+2022}. Using tuneable $\beta$-cooling prescriptions, \citet{Sudarshan_Penzlin_Cool+2022, WangCoolA+2022, WangCoolB+2022} find that slower cooling rates can result in drastic changes in disk response, which can result in the loss of accretion variability, depending on the adopted viscosity prescription.

 Overall, numerical simulations have shown that binaries can accrete at a high rate and can generate time variable, quasi-periodic accretion rate modulations, which can be translated into observed luminosity modulations. The shape of these time-series modulations and the dominant variability timescales depend on binary and disk properties. In some cases, periodic variability may be lost altogether. To prepare to interface with observations of accreting binaries, we require a more complete study of binary accretion. Specifically, numerical studies need to provide output appropriate for generating synthetic observations. At minimum, this should include 3D and more sophisticated treatment of the energy equation to allow for accurate inclusion of the evolution of and emission from optically thin regions such as may be found in the circumbinary disk cavity (see Figure~\ref{Fig:2DDens}). This must also be carried out for a range of binary and disk parameters.

\noindent
\begin{center}
{\textbf{ \em Binary Tidal Disruption}}
\end{center}
A subset of hydrodynamical variability signatures caused by binary accretion could come from the tidal disruption of a star by one of the SMBHs in a compact, sub-parsec separation binary. Such tidal disruption events (TDEs) might be expected if stellar hardening is responsible for SMBHB orbital evolution. The stellar disruption could serve as a fuel source for a circumbinary disk \citep{Coughlin_CBDfromTDE+2017}, or fall directly onto one of the black holes resulting in a TDE lightcurve that can be dramatically altered by the second black hole, for short orbital period SMBHBs. Large dips and excesses, or possibly periodic Doppler boosting (discussed below), on top of a decaying TDE lightcurve could help to identify the SMBHB \citep{LiuChen_TDESMBHB:2009, Ricarte+2016, HayasakiLoeb_TDESMBHB:2016, Coughlin_TDESMBHB+2017}. 

The probability of detecting such an event and the effect of the SMBHB on the TDE lightcurve is dependent on the binary separation \citep{Coughlin_TDEBin_review+2019}. At large separations, when one SMBH first strongly perturbs the stellar distribution bound to the other SMBH, the rate of TDEs can grow by 2-3 orders of magnitude over the single SMBH rate \citep[see][and refernce therein]{Chen_TDESMBHB+2011, LezhninVasiliev:2019}. Apart from indirectly inferring a high TDE rate in a single galaxy type (\eg, E+A galaxies \citealt{Arcavi+2014, French+2016}), these disruptions by widely separated SMBHBs likely look no different than a TDE onto a single SMBH. Anyway, after $\mathcal{O}$(Myr) the reservoir of available stars is quickly depleted and the TDE rate drops below the equivalent single SMBH rate \citep{Coughlin_TDEBin_review+2019}. The rate can again increase by a factor of few to ten if the binary reaches $\lesssim$milli-parsec separations \citep{Coughlin_TDESMBHB+2017, Darbha+2018}. At this stage, the binary orbital period becomes of order the stellar debris fallback time, allowing manifestation of the possibly identifiable binary signatures listed above. However, the enhanced TDE rate here may be mitigated by a shorter window of detectability set by the $\mathcal{O}$($10^3-10^9$)yr lifetimes for SMBHBs with $M=10^8 - 10^6 \Msun$, respectively. Candidate TDEs in binary systems have been proposed in the literature \citep{LiuKomossa_TDESMBHBcand+2014, CoughlinArmitage:2018, Shu_TDESBHB+2020}, each, however, with competing explanations \citep[\eg,][]{Grupe+2015, Leloudas+2016, Margalit+2018, Coughlin_DblStarTDE+2018}.

\subsubsection{\large  Orbital Doppler Boost and Binary Self-Lensing}
\label{S:DopLens}
Even in the case that accretion rate variations are not excited by the binary (\eg, $q \lesssim 0.05$) or are not translated into luminosity variations, periodic variability is expected to occur due to relativistic, observer-dependent effects when emitting gas is bound to one or both SMBHs in the binary. These include the relativistic Doppler boost caused by line-of-sight components of the orbital velocity, and binary-self-lensing, i.e., the periodic gravitational lensing of light from one SMBH's accretion disk by the other SMBH.

\noindent
\begin{center}
{\textbf{ \em Orbital Doppler Boost}}
\end{center}
The Doppler boost results in the luminosity of a source appearing brighter and bluer (or dimmer and redder), when the line-of-sight velocity is pointed towards (or away) from an observer. We briefly derive the expected Doppler boost variability \citep[see also][for a similar derivation]{Charisi+2022}. For a single, isotropically emitting source, the observed and emitted (primed frame) specific fluxes are related by
\begin{equation}
  F_{\nu}= D^{3-\alpha_{\nu}} F'_{\nu}, 
\end{equation}
where we have assumed additionally that the emitted spectrum of the source is a power law, $F_{\nu}\propto \nu^{\alpha_{\nu}}$, over the observing band with log slope $\alpha_{\nu} = \ln{F_{\nu}}/\ln\nu$.

The Doppler factor $D$ is the ratio of observed and emitted frequencies,
\begin{equation}
  D \equiv \frac{\nu}{\nu'} = \frac{1}{\gamma \left( 1 - \Vbet \cdot \hatn \right)},
\end{equation}
and depends on the Lorenz factor $\gamma$ and the line-of-sight velocity of the source $\Vbet \cdot \hatn$ in units of the speed of light.

Assuming a circular binary orbit with inclination $I$ measured from face-on, and expanding to first order in the orbital velocity $\beta_{\orb}$, the observed flux is,
\begin{equation}
F_{\nu} = (3-\alpha_{\nu}) \left[1 + \beta_{\orb} \cos{\left(\frac{2\pi}{P} t\right)} \sin{I} \right] F'_{\nu},
\label{Eq:DopFobs}
\end{equation}
which is a sinusoid with the same period as the orbit, $P$, and amplitude,
\begin{equation}
A_{\nu} = (3-\alpha_{\nu}) \beta_{\orb}  \sin{I}.
\label{Eq:DopAmp}
\end{equation}

\begin{figure}[ht]
\begin{center}$
\begin{array}{c}
\hspace{-10pt}
\includegraphics[width=\textwidth]{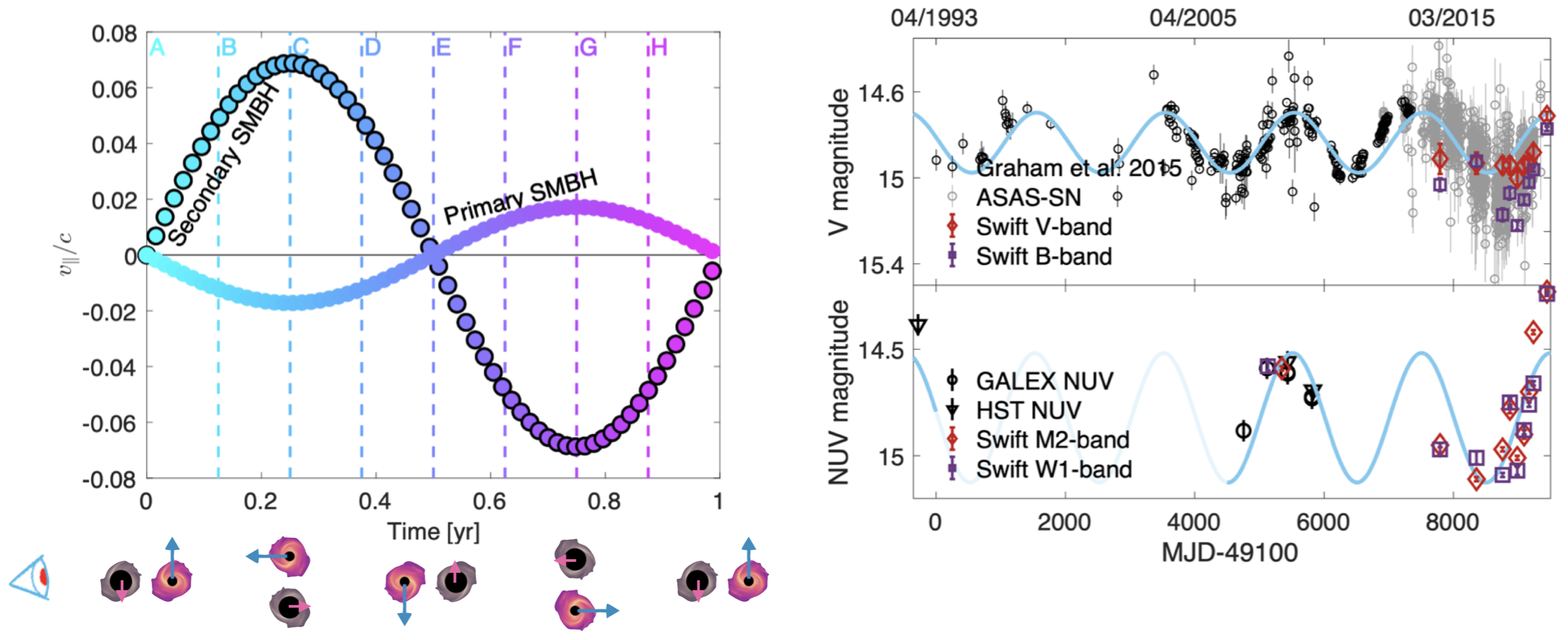} 
\end{array}$
\caption{\emph{Left}: Line-of-sight velocity for the primary and secondary SMBH in a binary with total mass $M=10^9M_{\odot}$, mass ratio $q=1/4$, period $P=1$\,yr and inclination $I=60^{\circ}$. The orientation of the binary with respect to the observer at various phases of the orbit is shown at the bottom of this figure. We have highlighted the mini-disk of the secondary SMBH, since it is typically more luminous and moves faster, and thus dominates the variability. Adapted from \cite{Charisi+2022}. \emph{Right}: The optical (top) and UV (bottom) lightcurve of PG 1302-102. The blue lines show the Doppler model given the optical and UV spectral indices. From \cite{ChengCheng_PG1302+2020}.
}
\label{Fig:Doppler_Boost}
\end{center}
\end{figure}

Writing $\beta_{\orb}$ for the secondary SMBH in the binary and in terms of number of Schwarzschild radii $N_a \equiv a/(2GM/c^2)$, $\beta_{\orb} =  \left[(1+q)^2 2 N_a\right]^{-1/2}$, which for SMBHBs at sub-parsec separations, corresponds to a few $\%$ to a $\sim 10\%$ amplitude modulation in the lightcurve (see left panel of Figure~\ref{Fig:Doppler_Boost}), which is easily observable with ground-based photometry, given that other sources of variability do not swamp the signal (see \S\ref{SubS:PLCsearches}). In the above, we assume one emitting source \eg, one of the mini-disks dominates the variability, typically the mini-disk of the secondary which also moves at higher velocities. If both mini-disks are Doppler boosted, since the two SMBHs are moving in opposite directions, the motion of one will lead to an increase in luminosity, while, at the same time, the motion of the other will have a dimming effect, diminishing the overall amplitude of variability.

Beyond motivating the observability of the orbital Doppler boost, Eq. (\ref{Eq:DopAmp}) offers a first consistency check for vetting this scenario. The photometric lightcurve of a periodic candidate provides the amplitude $A_{\nu}$ and period $P$, while a spectrum of the source provides the spectral index $\alpha_{\nu}$. An independent measurement of the total binary mass can often be obtained (though with large uncertainty) from the broad emission lines \citep{Shen_Greene_2008} or galaxy scaling relations \citep{Peterson+2014,ShenGreene+2015}. A combination of $P$ and $M$ give $\beta_{\orb}$ up to the mass ratio, and so the left and right-hand sides of Eq. (\ref{Eq:DopAmp}) are specified up to the binary mass ratio and inclination. 
Hence, an observed periodic lightcurve with a central mass measurement and spectrum can be used to rule out the Doppler hypothesis, or constrain the range of viable mass ratios and binary inclinations \citep{PG1302Nature:2015b, Charisi+2018}.

A more powerful consistency check comes with lightcurves in more than one observing band, \eg, centered at $\nu_1$ and $\nu_2$. By Eq. (\ref{Eq:DopFobs}), both lightcurves will have the same shape and period, but their modulation amplitudes are related by \citep{PG1302Nature:2015b},
\begin{equation}
\frac{A_{\nu_1}}{A_{\nu_2} } = \frac{(3-\alpha_{\nu_1})}{(3-\alpha_{\nu_2})}.
\end{equation}
Hence, lightcurves and spectra in two observing bands provide a parameterless consistency check for the Doppler boost model. 
This multi-wavelength Doppler boost test was successfully carried out in the optical and UV to describe the sinusoidal oscillations in the quasar PG 1302 \citep{Graham+2015a, PG1302Nature:2015b,ChengCheng_PG1302+2020},  
which is currently still being monitored to determine if the periodicity is persistent and preferred to quasar noise \citep{Liu_PG1302+2018, ZhuThrane:2020}. In the right panel of Figure~\ref{Fig:Doppler_Boost}, we show the multi-wavelength lightcurve of PG 1302 along with the Doppler boost model fit. The multi-wavelength Doppler signature was subsequently tested in 68 binary candidates from \cite{Graham+2015b,Charisi+2016} with optical and UV lightcurves finding that $\sim$1/3 are consistent with the model prediction \citep{Charisi+2018}. 
However, because quasars have wavelength-dependent variability  \citep{ChengCheng_UVopt_Cor:2020}, 
the signature can also be observed by chance in a control sample of non-periodic quasars at a rate of $20-40\%$, depending on the quality of the data \citep{Charisi+2018}. Therefore, it is challenging to distinguish this signature from intrinsic quasar variability, especially if the available multi-wavelength data are sparse.

 Note that if the source consists of not one isotropically emitting blob (\eg, the mini-disk bound to one of the SMBHs in the above derivation), but many such blobs moving with line-of sight speed $\Vbet \cdot \hatn$, then we must sum the specific intensity per unit length along the line of blobs, introducing an extra factor of $D$ from the isotropic case. This latter situation could describe a jet, for example. Conceptually, the line of emitting blobs is length contracted along the direction of motion, decreasing the flux by a factor of $D$,
 \begin{equation}
  F^{\mathrm{jet}}_{\nu}= D^{2-\alpha_{\nu}} F^{\mathrm{jet}'}_{\nu}, 
\end{equation}
where the time-dependent line-of-sight velocity will also be different from the single blob (disk case). If we assume that the line of emitting blobs is indeed a jet attached to one of the binary components, then the first-order effect is for the binary orbital motion to change the jet to line-of-sight angle, causing periodic brightness modulations with a different amplitude than in the disk case. This distinction could help discern variability with a jet origin from that with a disk origin. \citet{ONiell_ReadBland+2022} use the jet-Doppler model to explain the variability of the binary candidate PKS 2131−021 \citep{Ren+PKS0153_2021}.

Finally, for demonstrative purposes, we above assumed a circular orbit and that all of the light came from the secondary SMBH. Both of these assumptions can be relaxed resulting in a deviation of the lightcurve from a sinusoid \citep{SpikeyHu+2020,Charisi+2022}. Deviations from sinusoidal variability can 
also occur if the intrinsic source luminosity is not constant \citep{Charisi+2022}. Note that here, we also did not include percent-level effects due to finite-light travel time (R{\o}mer delay), or general relativistic pericenter precession, which can become important for highly eccentric orbits. All such terms can in principle be added to the Doppler model.

\noindent
\begin{center}
{\textbf{ \em Binary Self-Lensing}}
\end{center}

If the binary is inclined sufficiently close to the line-of-sight, then the light from the accretion disk around one SMBH will be lensed once per orbit by the other SMBH in the binary. The characteristic scales of the problem relevant for lensing are (1) the Einstein angle at conjunction,
\begin{equation}    
    \theta_E = \sqrt{\frac{2 R_{Sl} a}{D_A^2}},
\end{equation}
for Schwarzschild radius of the lens $R_{Sl} \equiv 2GM_{\rm{lns}}/c^2$ with $M_{\rm{lns}}$ the mass of the lensing SMBH, (2) the binary angular scale, $\theta_{\bin} \equiv a/D_A$, for a binary with orbital separation $a$ at angular diameter distance $D_A$ and, (3) a minimum size scale of emission at a given wavelength, $\theta_{\min} \equiv 6 G M_{\rm{src}}/(D_A c^2)$, approximated here by the innermost stable circular orbit (ISCO) of the SMBH which is being lensed, with mass $M_{\rm{src}}$. With these scales, we follow \citet{DODi:2018}; \citet{DODi:2020, DOLoebGWlens:2021} to derive the probability for a significant lensing event. Assuming that we observe the binary for at least one orbit, the probability is found from counting all ring orientations on the sphere for which the source passes within the Einstein radius of the lens at conjunction. The result approximates to
\begin{equation}
    \mathcal{P} \approx \frac{\theta_{E}}{\theta_\bin} = \sqrt{\frac{2 R_{Sl}}{a}} \approx 0.1 \left(\frac{a}{200 R_{Sl}} \right)^{-1/2}.
\end{equation}

The duration of the lensing flare will be
\begin{equation}
    \mathcal{T} \approx \frac{\theta_{E}}{\theta_\bin} P = \mathcal{P} P
    \approx 0.03 \mathrm{yr} \left(\frac{1+q}{2} \right)^{-3/2}  \left(\frac{a}{200 R_{Sl}} \right) \left(\frac{M}{10^8 \Msun} \right),
\end{equation}
which is one tenth the duration of the orbit in this example.

The magnification of a point source by a point mass lens is,
\begin{equation}
    \mathcal{M} = \frac{u^2 + 2 }{u\sqrt{u^2+4}},
\end{equation}
where $u$ is the sky-projected separation between source and lens in units of $\theta_E$. 
A significant lensing event is defined to occur at $u=1$, where $\mathcal{M} \approx 1.34$. For close encounters of the projected point-source and lens positions, $ u \ll 1$, and the magnification limits to $\mathcal{M}\rightarrow 1/u$. Hence the maximum magnification can be estimated as $1/u_{\min} = \theta_E/\theta_{\min}$,
\begin{equation}
    \mathcal{M}_{\max} \approx \frac{\theta_E}{\theta_{\min}} \approx \sqrt{\frac{2}{9} \frac{1}{q^2}  \frac{a}{R_{Sl}} } \approx \frac{20}{3} q^{-1} \left(\frac{a}{200 R_{Sl}}\right)^{1/2},
\end{equation}
reaching a magnification of 10 for $q=2/3$.

\begin{figure}[ht]
\begin{center}$
\begin{array}{c c}
\includegraphics[scale=0.45]{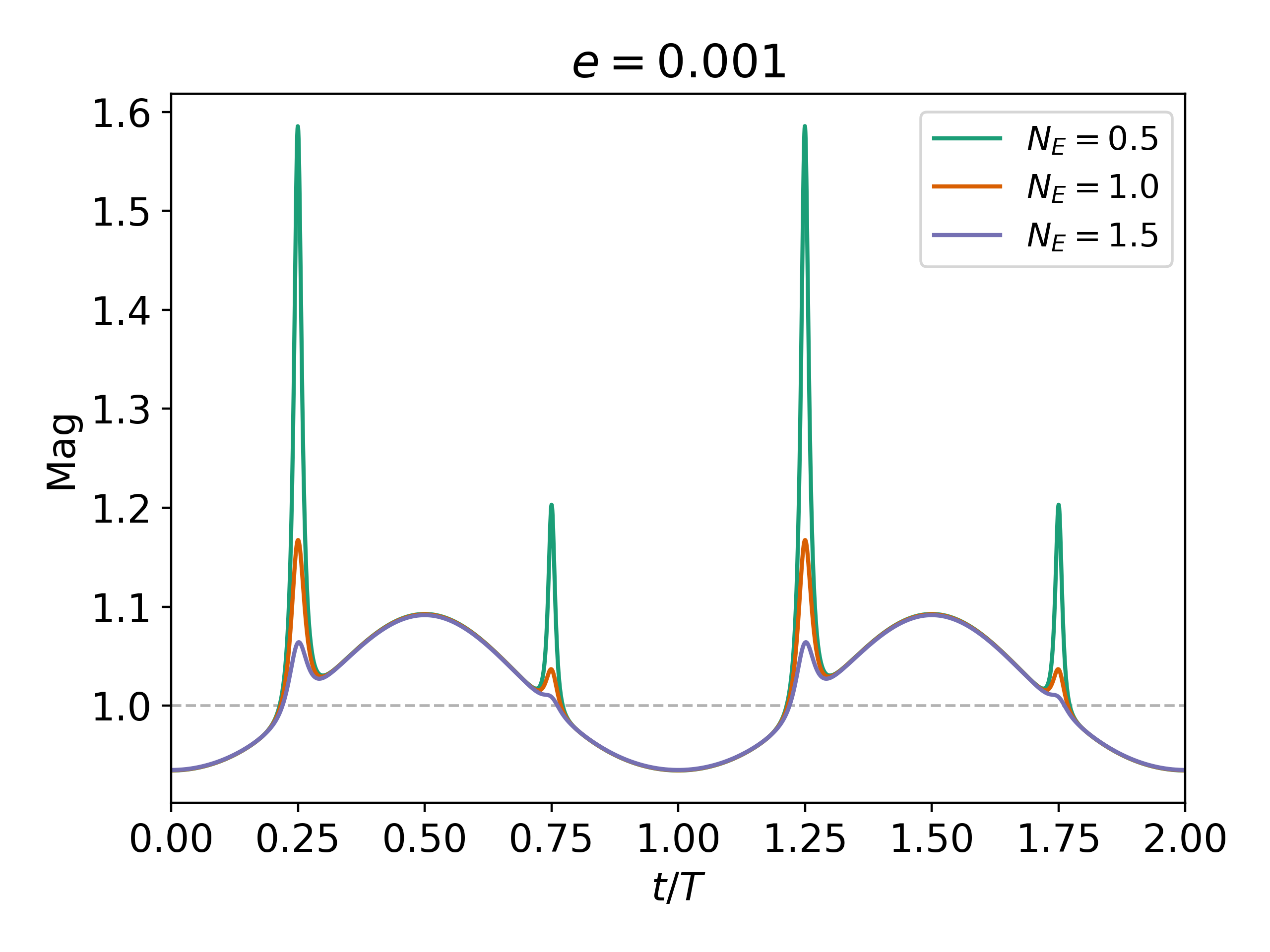} &
\includegraphics[scale=0.45]{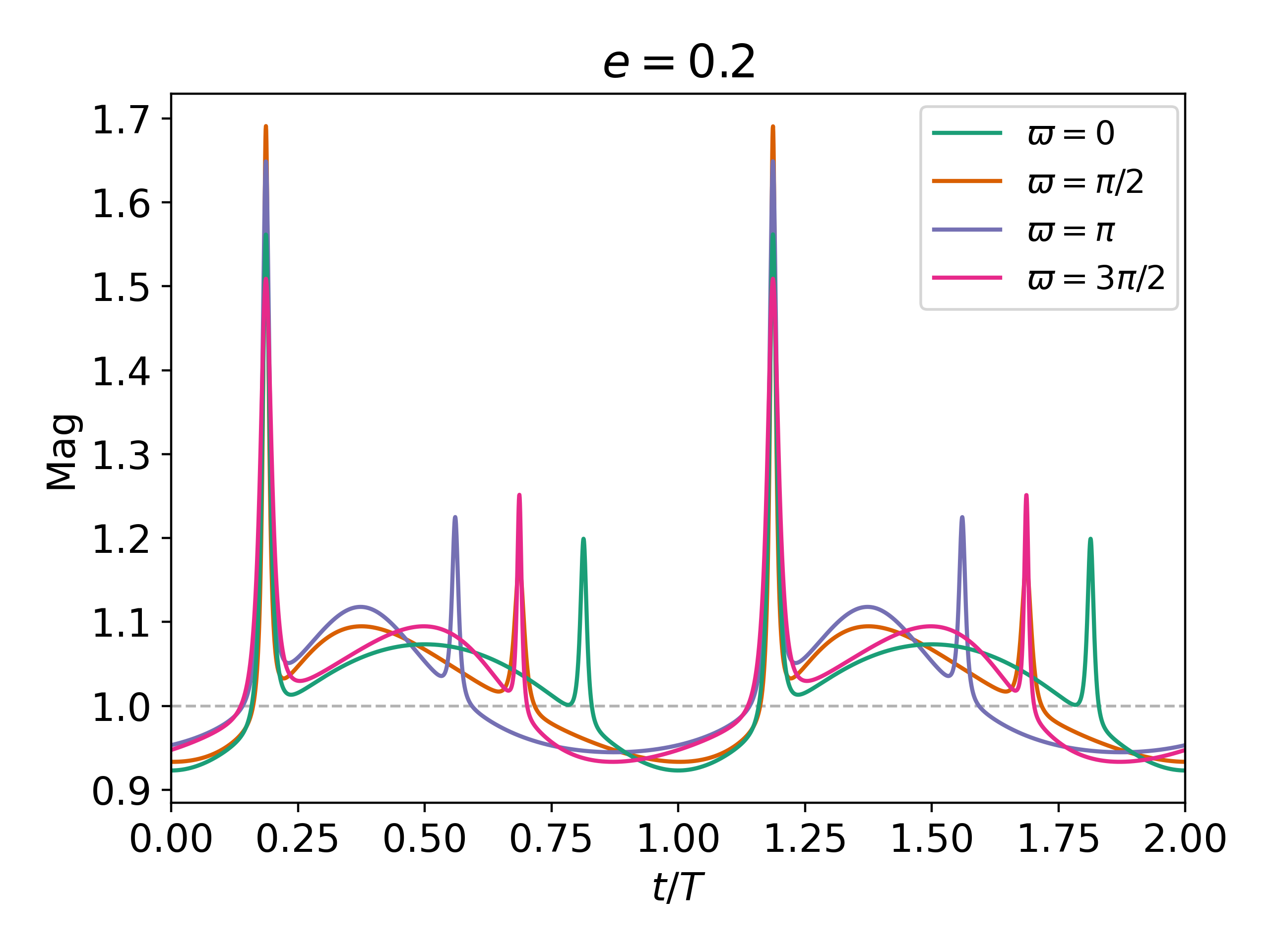} \\
\includegraphics[scale=0.45]{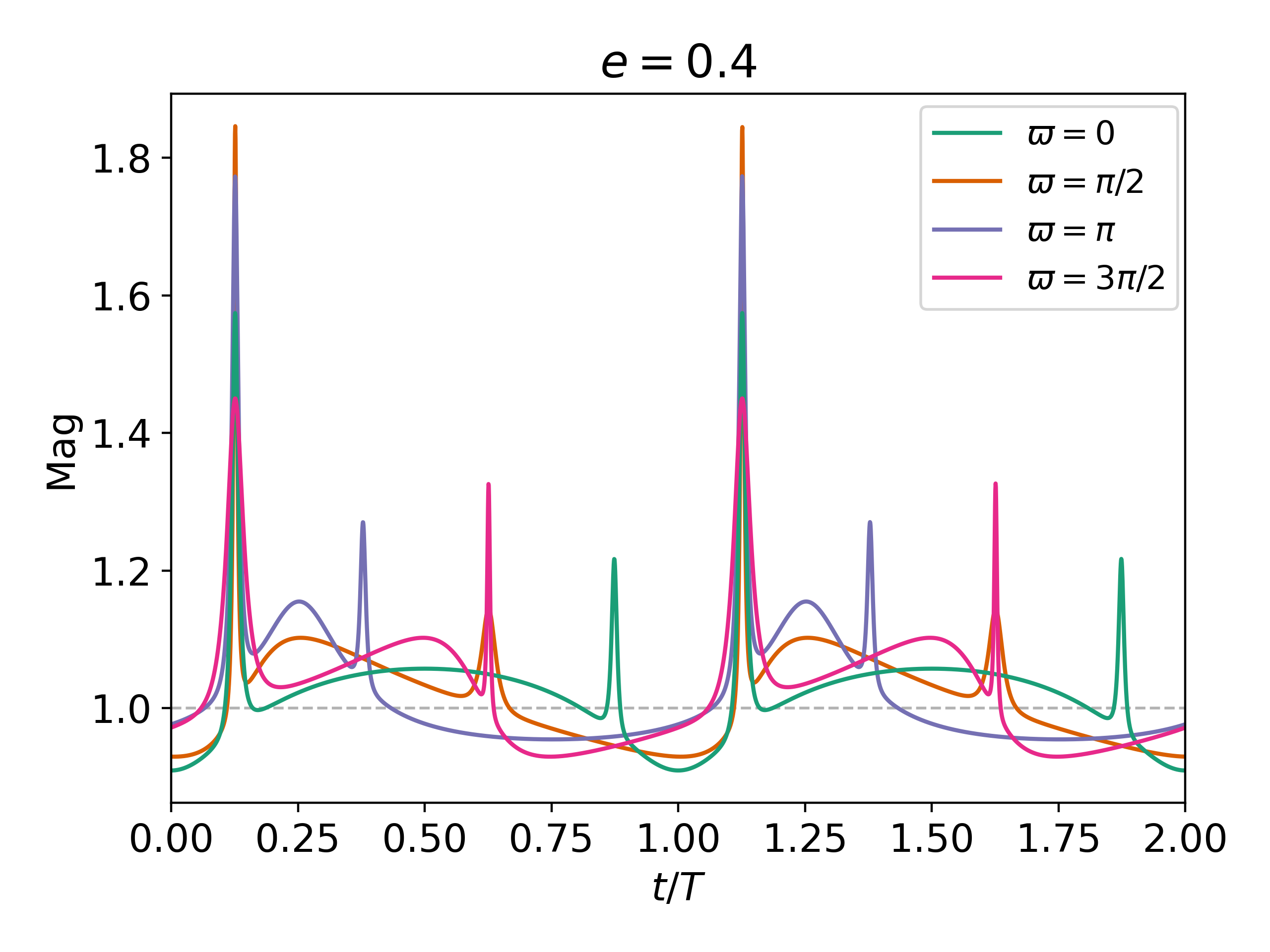} &
\includegraphics[scale=0.45]{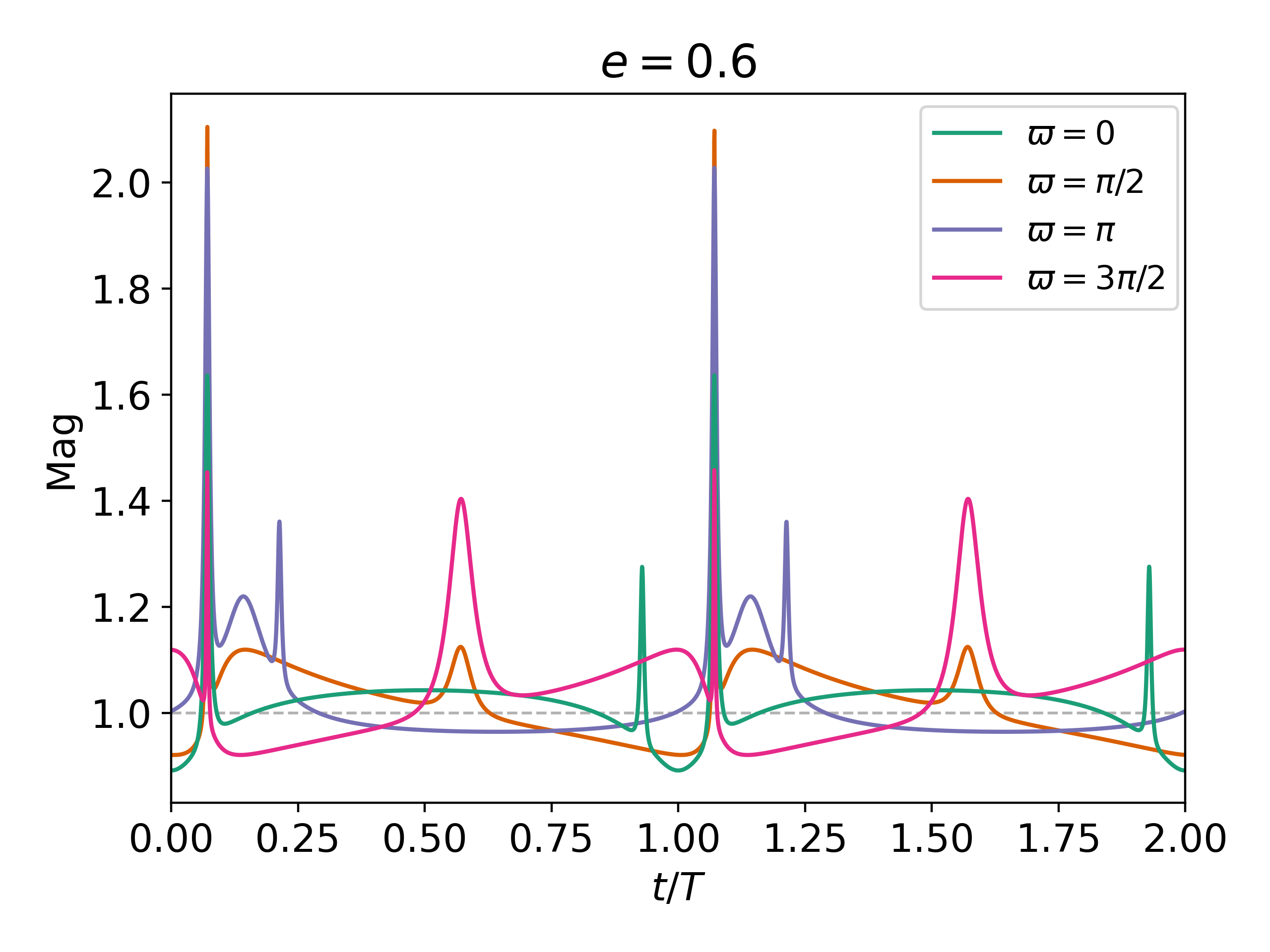} 
\end{array}$
\vspace{-5pt}
\caption{
Example Doppler+lensing lightcurves with parameters detailed in the text. The {\em top left panel} shows the change in lightcurve with inclination, for inclinations that cause a $N_E=0.5\times, 1.0\times$, and $1.5\times$ Einstein separation of source and lens at conjunction. The remaining panels show how the Doppler+lensing signature changes for eccentric orbits with different eccentricities and arguments of pericenter $\varpi$.
}
\label{Fig:DopLens}
\end{center}
\vspace{-10pt}
\end{figure}

Lightcurves of a steadily accreting source viewed at an angle where both Doppler boosting and lensing are prominent is shown in Figure~\ref{Fig:DopLens}. The lensing signature is unique; for a circular orbit (top left panel) it manifests as a symmetric flare occurring at the flux average of the sinusoidal Doppler modulation -- for eccentric orbits, the symmetric flares persist, but occur with a different relative magnitude and width at different (predictable) phases of a non-sinusoidal Doppler oscillation. The top right and bottom panels of Figure~\ref{Fig:DopLens} show how the lightcurve changes with eccentricity $e$ and argument of pericenter $\varpi$.\footnote{The model has ten parameters. In addition to the specified values of $e$ and $\varpi$, the parameters chosen for Figure~\ref{Fig:DopLens} are $M=10^9 \Msun$, $q=0.3$, $P=5$~yrs, $\alpha_{\nu}=-2$, $f_L=0.5$, $N_E=0.5$, $v_z=0$, $t_0=0$, where $f_L$ is the ratio of light coming from the secondary compared to the primary, $v_z$ is the barycenter velocity of the source in the line-of-sight direction, and $t_0$ is the starting epoch. For these parameters $N_E=0.5$ corresponds to a binary inclination of $I=87.08^{\circ}$.} Note that for eccentric orbits, the orbit can experience pericenter precession and the timing of the lensing flares would drift. This precession could be due to general relativity or sufficient amounts of surrounding matter.

In the case of a point source, this flare will be achromatic.
When the emission region size is of order the Einstein radius or larger, finite-source-size effects cause the shape and magnification to become wavelength dependent. Specifically, the shape and magnification of the lightcurve will depend on the shape and size of the emitting region at a given wavelength. Examples and specific criteria for a steady-state accretion disk source are given in \cite{DODi:2018}. In more extreme situations, finite lensing could even probe horizon scale features of the accretion flow around the source SMBH \citep{JordyZ:2022, JordyZLetter:2022}. In addition to repeating lensing flares, matter in the accreting binary may also cause eclipses \citep{Ingram+2021}.

The best candidate for such a Doppler plus self-lensing SMBHB derives from the Kepler light-curve for the ``Spikey'' system, which is remarkably well fit by an eccentric SMBHB Doppler+lensing model \citep{SpikeyHu+2020}, albeit for only one flare on top of a sawtooth-shaped modulation. From observations of one putative lensing flare, however, an orbital period can be inferred allowing predictions for future flares and so a test of the SMBHB model, which could rule out or provide convincing evidence for a milli-parsec separation SMBHB (\textcolor{cyan}{D'Orazio et al. \textit{In prep.}}).

\clearpage

\subsubsection{\large  Reverberation}
\label{SubS:Reverb}
The above considers hydrodynamic, Doppler, and self-lensing sources of periodic variability in directly observed continuum emission. A more complete description must place this emission in the broader context of AGN. Specifically, the medium surrounding the central AGN power source can reprocess this emission via absorption and scattering processes. As we saw in \S\ref{SubS:BLRs}, broad-line reprocessing is one example of such reverberation of the central continuum emission.

\begin{wrapfigure}{r}{0.5\textwidth} 
\begin{center}$
\begin{array}{c}
\hspace{-5pt}
\includegraphics[width=0.5\textwidth]{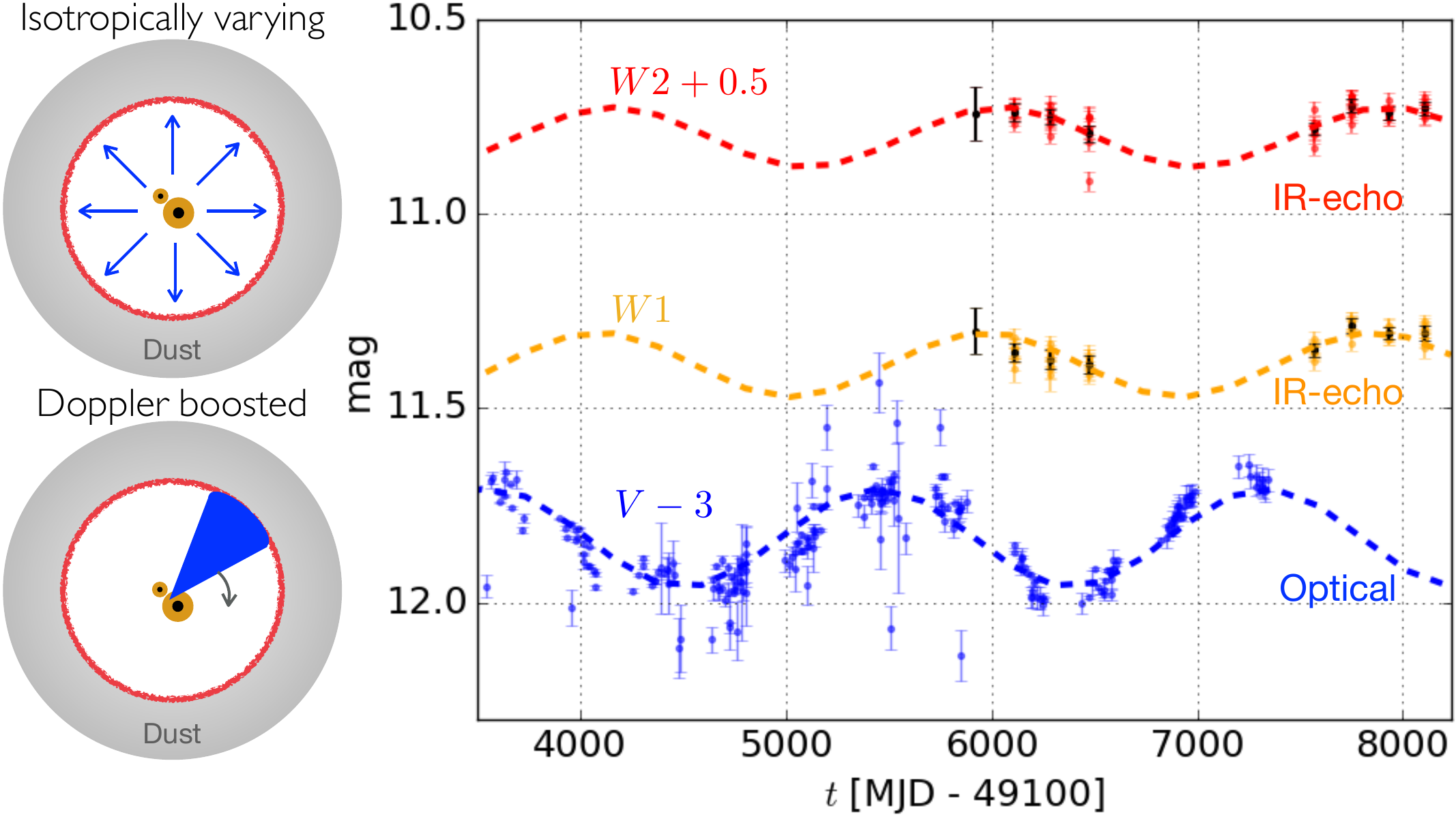} 
\end{array}$
\caption{\emph{Left:} Schematics representing dust tori (grey) illuminated by optical/UV emission (blue) from a central accreting binary, thereby heating the dust (red rings) causing it to echo the continuum emission in infrared emission. Isotropically varying and Doppler boosted emission will illuminate the dust differently resulting in different echoed IR emission (\S\ref{SubS:Reverb}). \emph{Right:} Optical (V-band, blue) and IR (WISE W1- and W2-bands, orange and red) emission from SMBHB Doppler boost candidate PG 1302-102 showing phase-lagged and diminished amplitude variability in the IR \citep{Jun_IRLag+2015, DZ_Lighthouse:2017}, dashed lines are sinusoidal fits. 
}
\label{Fig:DustIR}
\end{center}
\end{wrapfigure}

Beyond the broad line region, optical to UV radiation can be absorbed by surrounding dust structures. This results in an echoed infrared (IR) continuum with the same periodicity as the optical/UV, but with a time delay and a diminished amplitude of variability. While any optical/UV continuum variations can be echoed by dust, \citet{DZ_Lighthouse:2017} predict time delays and variability amplitude ratios between IR and optical/UV lightcurves that differ for reverberation of spatially isotropic illumination (\eg, periodic accretion variability of \S\ref{SubS:HydroVar}) vs. anisotropic illumination via the Doppler Boost (see Figure \ref{Fig:DustIR}). Three main differences in particular are: (1) periodically variable IR lightcurves caused by the Doppler boost have an extra quarter cycle phase lag compared to the isotropic reverberation case, for the same reverberating dust structure. (2) In the limit of small light travel time across the dust structure compared to the variability period, isotropically illuminated dust structures echo a maximum variability amplitude equal to the continuum level, whereas the echoed Doppler boosted variability amplitude drops to zero, completely washing out the IR signal (see Figure 3 of \citealt{DZ_Lighthouse:2017}). Finally, (3) dust illumination by Doppler boosted emission allows for a new possibility of ``orphan IR periodicity''; for example, a near face-on binary does not produce any Doppler boosted optical/UV periodicity, since the line-of-sight velocity is small,  but dust at different inclinations to the binary plane than the observer does see the anisotropic, time variable Doppler boosted illumination. Hence, constant continuum UV/optical emission would be accompanied by periodic variability in the echoed IR emission. This signature is yet to be observed. Using WISE IR data, \citet{Jun_IRLag+2015, DZ_Lighthouse:2017} utilize these IR echo models to further vet the Doppler-boost hypothesis for the SMBHB candidate PG 1302 (see Figure \ref{Fig:DustIR}).

Scattering of the direct continuum emission from each accreting SMBH in the binary could lead to a time-changing polarization degree and polarization angle of the observed emission. \citet{DottiPol+2022} consider scattering of light emitted from one of the binary components by a surrounding circumbinary ring. For fraction $s$ of light scattered by the ring, \citet{DottiPol+2022} predict $0.2s$-level periodic variability in the polarization degree, modulated at the binary orbital period, with minimum polarization degree coincident with the maximum of the continuum flux. Periodic variability in the polarization angle occurs at the $1\times$ or $2\times$ the binary orbital period with typical amplitude of variation at the $1^{\circ}$ level. Both are at the limits of detection via modern polarimetry of AGN \citep[see][and references therein]{DottiPol+2022}.

While each of the above reprocessing phenomena occur also for continuum emission from a single SMBH, the time changing distance from the binary to the reprocessing region caused by orbital motion, as well as temporal, and sometimes spatial, anisotropic illumination could imprint identifiable features on the reprocessed emission.
Hence, these reprocessed signatures offer an extra handle on distinguishing the causes of periodic variability of AGN lightcurves and so for testing the SMBHB hypothesis.

\subsubsection{\large  Periodicity Searches and Sources of Noise and Confusion}
\label{SubS:PLCsearches}

The search for quasars with periodic variability has seen rapid developments in the last few years due to the advent of time-domain surveys, which systematically scan the sky and record the brightness evolution for millions of sources. These surveys have provided large samples of quasar lightcurves and allowed for the first systematic searches, which have returned $\sim$250 SMBHB candidates. 

More specifically, \citet{Graham+2015b} analyzed a sample $\sim$245,000 quasars from the Catalina Real-time Transient Survey (CRTS) and identified significant evidence for periodicity in 111 quasars. \citet{Charisi+2016} performed a systematic search in a sample of $\sim$35,000 quasars from the Palomar Transient Factory (PTF) and detected 33 candidates. One additional candidate was found in a small sample of 9000 quasars from the Panoramic Survey Telescope and Rapid Response System Medium Deep survey (Pan-STARRS MDS; \citealt{LiuGezari+2019}), while five more candidates were discovered in an even smaller sample of 625 AGN from the Dark Energy Survey (DES; \citealt{Chen_DES_PLCs+2020}). A larger sample of 123 candidates was detected recently from the analysis of a sample of 143,700 quasars from the Zwicky Transient Facility (ZTF; \citealt{Chen_ZTF_PLCs+2022}). 

The above candidates have periods from a few hundred days to a few years, with separations ranging from a few to a few hundred milli-parsecs. The magenta dots \citep{Graham+2015b}, orange x's \citep{Charisi+2016}, black x \citep{LiuGezari+2019}, black +'s \citep{Chen_DES_PLCs+2020}, and chartreuse stars \citep{Chen_ZTF_PLCs+2022} in Figure~\ref{Fig:BHBDemo} denote photometric variability candidates from CRTS, PTF, Pan-STARRS MDS, DES and ZTF surveys, respectively. The range of searched (and thus detected) periods is dictated by the data. The baseline of the lightcurve (i.e. the maximum interval of observations) determines the maximum period, with most searches requiring that at least 1.5 cycles of periodicity are present within the available baseline. The minimum allowed period is determined by the survey cadence, i.e. how often observations are taken, but most studies have limited the searches to periods of more than a month to avoid unmodeled photometric effects mainly from the lunar cycle. It is also worth pointing out that even though the above studies employed distinct search methods, the rate of periodicity detection is within an order of magnitude, i.e., $10^{-3} - 10^{-4}$.

Identifying periodicity in quasars is extremely challenging and the above samples likely contain a large number of false detections. The main reason is that quasars have stochastic variability which can both mimic periodicity, introducing false positives, and hinder the detection of real periodic signals \citep{Vaughan2016, WittCharisi+2022}. This effect is pronounced for relatively long periods for which we cannot observe many cycles within the available baselines. We emphasize that despite the common misconception that the above samples of periodic candidates contain false positives because the stochasticity of the noise was not included in the statistical analysis, all of the above systematic searches assumed that the quasar noise is described by a Damped Random Walk (DRW) model, which is currently the most successful description of quasar variability \citep{MacLeod2010,Kozlowski2010}. In particular, \citet{Graham+2015b} used a combination of wavelets and auto-correlation function to select periodic candidates. Then for each observed lightcurve, they simulated a DRW lightcurve with properties like the observed lightcurve (sampling, gaps, photometric errors) and showed that the mock population cannot produce candidates that pass the periodicity detection threshold. \citet{Charisi+2016} used the Lomb-Scargle periodogram to select quasars with periodic variability, and assessed the periodicity significance with DRW simulations. For each lightcurve, they fit a DRW model and subsequently simulated 250,000 mock lightcurves with the best fit DRW parameters and properties like the observed to account for trial factors from searching in a large sample and over multiple frequency bins. They assessed a false alarm probability for each source by comparing the periodogram peak of the observed lightcurve with the distribution of peaks from the simulations. \citet{Chen_DES_PLCs+2020} followed a very similar approach to \citet{Charisi+2016}, but generated 50,000 mock lightcurves for each observed lightcurve, since the search sample was smaller. \citet{LiuGezari+2019} pre-selected periodic candidates using the Lomb-Scargle periodogram under the assumption of white noise and down-selected candidates by performing a maximum likelihood selection between a DRW and a DRW+periodic model. Finally, \citet{Chen_ZTF_PLCs+2022} followed a similar approach to \citet{LiuGezari+2019}.

\cite{Sesana+PLC_PTA+2018} also demonstrated that not all candidates can be genuine binaries by extrapolating the population of binaries given the candidates identified in the CRTS \citep{Graham+2015b} and PTF \citep{Charisi+2016} samples. They concluded that the expected GW background from the inferred population is in moderate tension with upper limits from PTAs, which further indicates that the samples contain false positives. Using similar population arguments, \citet{Holgado+2018} estimated the maximum fraction of blazars (a sub-class of quasars in which the relativistic jet is pointed towards our line-of-sight) that can be associated with short-period binaries and suggested that the binary scenario is not a good interpretation for the year-long quasi-periodicities observed in blazars.

Given the contamination with false detections, follow-up efforts have focused on collecting multi-wavelength data in order to find converging evidence that could signify the presence of a binary in candidate quasars, covering the entire array of potential signatures described above, such as (1) variability in the broad emission lines \citep[\S\ref{SubS:BLRs};][]{Song_PG1302_BLR+2021, Ji_CBD_BLR+2021, NguyenBogIII:2020} (2) spectral X-ray signatures \citep[\S\ref{SubS:CBDspectra};][]{Saade_Chandra:2020}, (3) periodicity with multiple periodic components and a characteristic frequency pattern \citep[\S\ref{SubS:HydroVar};][]{PG1302MNRAS:2015a, Charisi+2015}, (4) multi-wavelength Doppler boost or self-lensing \citep[\S\ref{S:DopLens};][]{PG1302Nature:2015b, Charisi+2018, ChengCheng_PG1302+2020}, (5) the IR dust reverberation signature \citep[\S\ref{SubS:Reverb};][]{Jun_IRLag+2015,DZ_Lighthouse:2017}, and (6) the imprint of binaries on radio jets \citep[\S\ref{S:JetMorph};][]{Kun+2015:PG1302, Mohan_PG1302+2016, Qian_PG1302Jet+2018}. A subset of the candidates show one or more of the above promising signatures and warrant further monitoring.
However, not all of these signatures are unique to binaries, nor is it always clear when multiple effects should occur at the same time, and thus they cannot be easily distinguished from the intrinsic variability of typical single-SMBH quasars \eg, \cite{Charisi+2018}.

As is apparent from the above, the optical band dominates the discovery space in time-domain searches, due to the abundance of photometric data produced from ground-based surveys primarily during the last decade. Time-domain surveys have operated across the electromagnetic spectrum for a while (e.g., UV/GALEX; \citealt{GALEX}, mid-IR/NEOWISE; \citealt{NEOWISE}, X-rays/Swift; \citealt{Swift}, eROSITA; \citealt{eROSITA}, gamma-rays/Fermi LAT; \citealt{Fermi_LAT}), but these surveys typically do not have the required combination of data quantity (as mentioned in \S\ref{S:popexp},
 SMBHBs are relatively rare and thus a large sample of quasars is necessary) and data quality (long baselines, with dense enough sampling) to allow for similar systematic searches. One exemption is the Swift BAT survey spanning 105 months, in which two independent systematic searches in a small sample of <1000 AGN have returned one additional candidate \citep{Swift_BAT_Serafinelli,Swift_BAT_Liu}. This rate is roughly consistent with the optical searches above.

Beyond the systematic searches, there have been multiple reports of serendipitous detections of quasar periodicity in multiple bands. In \S\ref{apendix:Phot}, we present a likely incomplete summary of these candidates.

 \vspace{0pt}
 \begin{center}
 \begin{tcolorbox}[width=0.98\textwidth,colback={white},title={{\bf Pros and Cons: Periodic Lightcurves}}, colbacktitle=gray,coltitle=black]    
 \textbf{Pros:}
 \begin{itemize}
     \item The expectation for periodic brightness modulation in binaries is simple and intuitive -- the two body problem is periodic. Many mechanisms can produce periodicity, as shown in a variety of analytical and numerical studies.
     \vspace{-5pt}
     \item Some signatures (\eg, the relativistic Doppler boost) are unique containing testable multi-wavelength predictions and possible reverberation signatures.
     \vspace{-5pt}
     \item This method has been applied in large samples of AGN from past or ongoing time-domain surveys revealing hundreds of promising candidates. 
      \vspace{-5pt}
     \item Upcoming time-domain surveys like LSST will provide even bigger samples, which are required for the detection of rare short-period binaries.
 \end{itemize}
 \textbf{Cons:}
 \vspace{-5pt}
 \begin{itemize}
     \item AGN variability is stochastic and can mimic periodicity. Therefore, the samples of candidates are likely contaminated with false detections.
     \vspace{-5pt}
     \item Not all signatures are unique, making them hard to distinguish from typical quasar variability, especially in the limit of sparse data.
     \vspace{-5pt}
     \item Building confidence in the candidates requires observations of many cycles of periodicity, which for periodicities of a few years requires time (decades).
     \vspace{-5pt}
     \item The multi-wavelength Doppler boost test requires spectra and multi-epoch photometry in two observing bands, while lensing requires relatively high cadence over long temporal baselines.
     \vspace{-5pt}
     \item Application of IR reverberation models require high-cadence IR all-sky surveys (\eg, to build a large sample and systematically search for binaries) or long-term follow-up of individual candidates. 
     \vspace{-5pt}
     \item Polarization signatures are at the limit of detection and can be confused by scattering from intervening gas in the AGN host galaxy.
     \vspace{-5pt}
     \item Models of binary accretion are still young.
 \end{itemize}
 \end{tcolorbox} 
 \end{center}

\subsection{\large  \large  Jet Morphology}
\label{S:JetMorph}
Searching for peculiar or spatially periodic features in AGN jets was one of the first methods proposed for finding SMBHBs \citep{BBR:1980}. The idea is that accretion onto the binary may result in a jet launched from one or both SMBHs. The binary orbit (or the binary coalescence) can be imprinted on the large-scale features of the jet (see left panel of Figure~\ref{Fig:JetCombined}).

\subsubsection{\large  Periodic Wiggles and Helicity} 
In one set of models, shocks at the jet base launch plasmoids that travel away from the central power source at relativistic velocities on ballistic trajectories. If the jet power source is stationary, these plasmoids will trace out a line in space, resulting in a straight collimated jet \citep{Rees_Jets:1978}. If the jet is precessing, the plasmoids will trace out a conical helix, whereas if the base of the jet is moving, e.g., due to the orbital motion in a binary system, the shape of the jet depends on the orientation of the jet axis with respect to the binary orbital plane, with the plasmoids tracing a cylindrical/conical helix if the jet axis is perpendicular/titled with respect to the orbital plane \citep{Rieger2004}. 

The theory describing the projected shapes of precessing jets was first laid out in \citet{Gower+1982}. In a precessing jet, the opening angle, $\Psi$, is given by the amplitude of the precession angle, while the wiggle wavelength is set by the precession period and sky projection. Jet precession can be observed both in a single-SMBH AGN and in a binary system. For a single SMBH, precession can be caused by Lense-Thirring precession of a disk misaligned with the SMBH spin \citep[see][and refernces therein]{Liska2018}.
In a binary, precession could be caused by geodetic precession of the SMBH spin which is misaligned with the binary orbital angular momentum \citep{BBR:1980, Roos:1988}, or torquing of a disk around the jet-launching SMBH, which is misaligned with the binary orbital plane. 
The former occurs on long timescales, upwards of thousands of years,
\begin{equation}
P_\mathrm{geop} \approx 9 \times 10^3   \mathrm{yr} \ \ q^{-1} \left(\frac{a}{10^{-2} \mathrm{pc}}\right)^{5/2}  \left(\frac{M}{10^8 \Msun}\right)^{−3/2} .
\label{Eq:GeoPrec}
\end{equation}
However, if the disk is torqued by a binary companion at distance $a$, then the precession timescale, $P_{\mathrm{prc}}$, of a disk around the secondary (primary), perturbed by the primary (secondary) is significantly shorter \citep{LaiPrec:2014},
\begin{equation}
\frac{P_{\mathrm{prc}}}{P_{\orb}} = -\frac{8}{3} \frac{1+q^s}{\sqrt{1+q^{-s}}} 
\left( \frac{a}{r_d} \right)^{3/2} (\cos{\chi})^{-1};  
\label{Eq:TrqPrec}
\end{equation}
where $s=1$ ($s=-1$) corresponds to the secondary (primary) disk perturbed by the primary (secondary), $\chi$ is the angle between the disk angular momentum and the binary orbital angular momentum, $r_d$ is the outer radius of the precessing disk. This calculation also assumes that the disk surface density falls as $1/r$.

Even if neither of the above precession mechanisms operate, the binary orbital velocity can generate a similar effect. If the jet launching SMBH has an orbital velocity due to its motion in a binary system, then the plasmoids are launched with direction and speed given by relativistic velocity addition of the jet velocity, $\mathbf{v}_{\mathrm{jet}}$, and the orbital velocity of the base of the jet, $\mathbf{v}_{\orb}$. Hence, the jet launching velocity is modulated in direction and magnitude over the course of an orbit. This also results in plasmoids tracing a cone (or a cylinder) with opening angle given by $\Psi \approx v_{\orb}/{v_\mathrm{tot}} \cos{\chi}$ \citep{Roos:1993}, but now the wiggle wavelengths are set by the orbital period. 

It is important to note that the observed and source frame precession periods can be very different in the case of large jet velocities directed near to the line of sight. For jet velocity $\beta_j\equiv v_j/c$, time-dependent angle to the line of sight $\phi_{j}(t)$, and duration in source frame of ($t_{\mathrm{src}} - t_0$) the observer and source frame times are related by,
\begin{equation}
    \Delta t_{\mathrm{obs}} = (1+z) \int^{t_{\mathrm{src}}}_{t_0}{\left[1 - \beta_j \cos{\phi_{j}(t')}\right] dt'}
\end{equation}
\citep[\eg,][]{Roland2008, Caproni2013}. For example, the observer- and source-frame jet-precession periods measured for BL Lacertae (2200+420; \citealt{Caproni2013}) are $12.11$~yr and $550$~yr, respectively (translating to an inferred source-frame binary orbital period of $335$~yr in that model). For the candidate in PKS 1830-211, they are $1.08$~yr and $30.8$~yr, respectively \citep[\S\ref{apendix:jets};][]{Nair2005}. This has the benefit of probing much longer orbital periods than available through photometric periodicity (\S\ref{SubS:PLCsearches}).

\begin{figure}[ht]
\vspace{-10pt}
\begin{center}
\hspace{-10pt}
\includegraphics[width=0.9\textwidth]{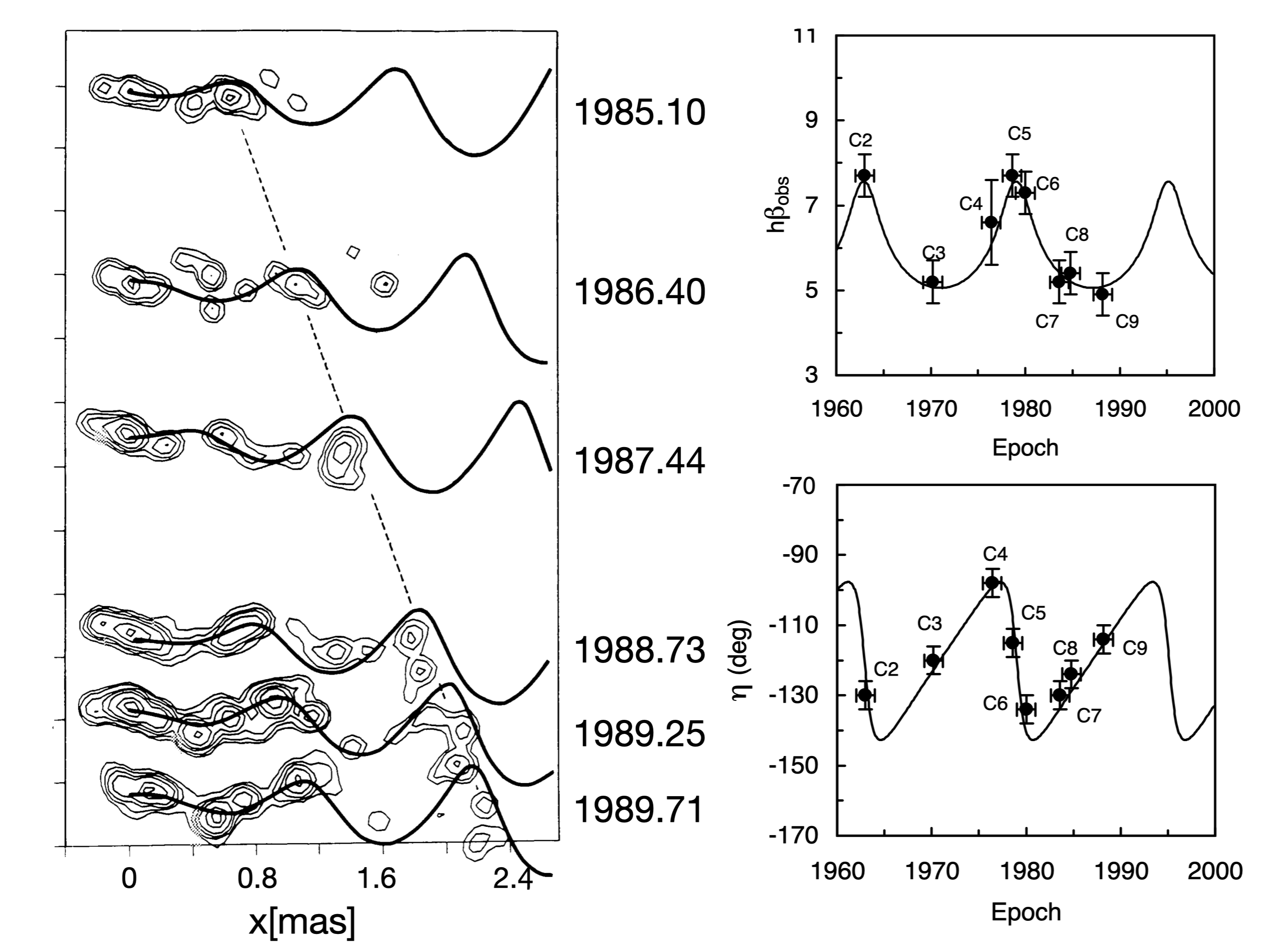}
\caption{\emph{Left}:
Example radio contours of the jet in quasar 1928+738 exhibiting ``wiggle'' and overlaid jet+SMBHB model, from \citet{Roos:1993}. \emph{Right}: Data for the positions of different jet components $\eta$ and corresponding observed proper motions in units of the speed of light (up to Hubble parameter $h$), $\beta_{obs}$, for the jet in quasar 3C 273. The periodic nature, over the considered epochs, can be fit by a model where the jet is attached to one of the SMBHs in a binary, from \citet{AbrRom:1999}. 
}
\label{Fig:JetCombined}
\end{center}
\end{figure}

Varying the viewing angle for either precession or orbital modulation models results in a zoo of possible observed jet shapes, which encode the periodic nature of the jet launching. In addition to the projected positions of jet plasmoids on the sky tracing out the jet shape, the above described ballistic models also predict the apparent velocity of each plasmoid and its Doppler factor. Fitting a jet model to the data requires tracking the apparent velocity and sky position of multiple plasmoids over time, in order to constrain the evolution of jet motion and disentangle the variety of parameters in the model (\eg, see right panels of Figure~\ref{Fig:JetCombined}).

Further modeling has included the impact on the jet morphology from the possible orbital decay due to GWs \citep{KulkLoeb:2016}, the possibility of emission from double (and possibly intersecting) jets \citep{Palenzuela_dbljet+2010, Gold:GRMHD_CBD:2014, Gold:GRMHD_CBDII:2014, BrightPasch_minidisks:2023}, and MHD simulations of a wobbling jet in an ambient medium, constraining jet precession angles for which successful propagation can occur \citep{Fendt_CurvedMHDJets:2022}.

Several candidates have been identified based on the above jet signatures. In \S\ref{apendix:jets}, we present a (likely incomplete) summary of the main candidates. The inferred binary properties from these candidates are plotted in Figure \ref{Fig:BHBDemo} as yellow `pluses' (we have only reported mean or typical inferred binary parameters and draw a down caret in the case of an upper limit). However, we emphasize that, like many of the aforementioned electromagnetic signatures of SMBHBs, jet wiggles are also not unique and may be produced by alternative mechanisms. For instance, instabilities such as the Kelvin-Helmholtz instability acting between the fast-moving jet material and its surroundings can produce distorted jets, mimicking the wiggles from a binary system \citep{LobanovZensus_3C273Jet:2001, 2011A&A...529A.113Z}. Long-term monitoring or higher-resolution imaging of the jets can help differentiate the possible scenarios.

Several studies have followed up candidates selected with a different method, e.g., quasars with periodic variability, to search for jet distortions. In some cases, these observations have found jet signatures consistent with the binary candidate properties and offer independent support for the binary nature of these sources. In \S\ref{apendix:jets:followup}, we briefly summarize these follow-up efforts and findings for the main candidates.


\subsubsection{\large  X's and Other Shapes} 
Another class of SMBHB jet morphology models have been devised to explain the 5-10$\%$ of radio galaxies that exhibit peculiar ``winged'' or X-shaped radio structures \citep{LeahyWilliams:1984, Yang+Xjet+2019}. Explanations relying on an SMBHB include mis-aligned jets attached to each SMBH in a binary or a merger-induced spin flip \citep{MerrittEkers:2002, GopalKrishna_XZjet+2003}. There exist, however, a range of suggested causes that do not invoke SMBHBs \citep[starting from][]{Rees_Jets:1978}, and including recent GRMHD simulations which show that a transient initial state in the jet launching process for a single SMBH can naturally generate X-shaped radio morphologies \citep{LalakosSasha+2022}.

 \vspace{0pt}
 \begin{center}
 \begin{tcolorbox}[width=0.98\textwidth,colback={white},title={{\bf Pros and Cons: Jet Morphology}}, colbacktitle=gray,coltitle=black]   
 \textbf{Pros:}
 \begin{itemize}
    \item Many jets with wiggles and and ``wing-'' or X-shapes have been observed resulting in many proposed binary candidates.    
    \vspace{-5pt}
    \item The periodic jet structures can sometimes probe long ($\sim100$~yr) orbital periods due to strong relativistic effects.
    \vspace{-5pt}
    \item Could learn from prototypical systems: galaxy 3C 75 with two jets, \citep[but separation of SMBHs is large,][]{3C75+1985}, or stellar-mass (micro-quasar) system with remarkable jet morphology, SS433 \citep[see][and refrences therein]{SS433_Blundell+2018}. 
 \end{itemize}
      \vspace{-5pt}
 \textbf{Cons:}
  \begin{itemize}
      \vspace{-5pt}
     \item These jet signatures are indirect and not unique and thus they can be confused with jet features generated by a single SMBH.
     \vspace{-5pt}
    \item Only $\sim10\%$ of AGN are radio loud.
 \vspace{-5pt} 
 \end{itemize}
 \end{tcolorbox} 
 \end{center}

\subsection{\large  \large  Direct Imaging and Orbital Tracking at Sub-Parsec Separations}
\label{S:DirImg}
Until now we have only considered indirect means for detecting sub-parsec separation SMBHBs. The reason, as discussed in \S\ref{S:popexp}, is the small angular diameter of the binary separation on the sky (Eq. \ref{Eq:thetBin}), of order $10 \mu$as or smaller.

\begin{figure}[ht]
\begin{center}$
\begin{array}{c}
\hspace{-10pt}
\includegraphics[scale=0.4]{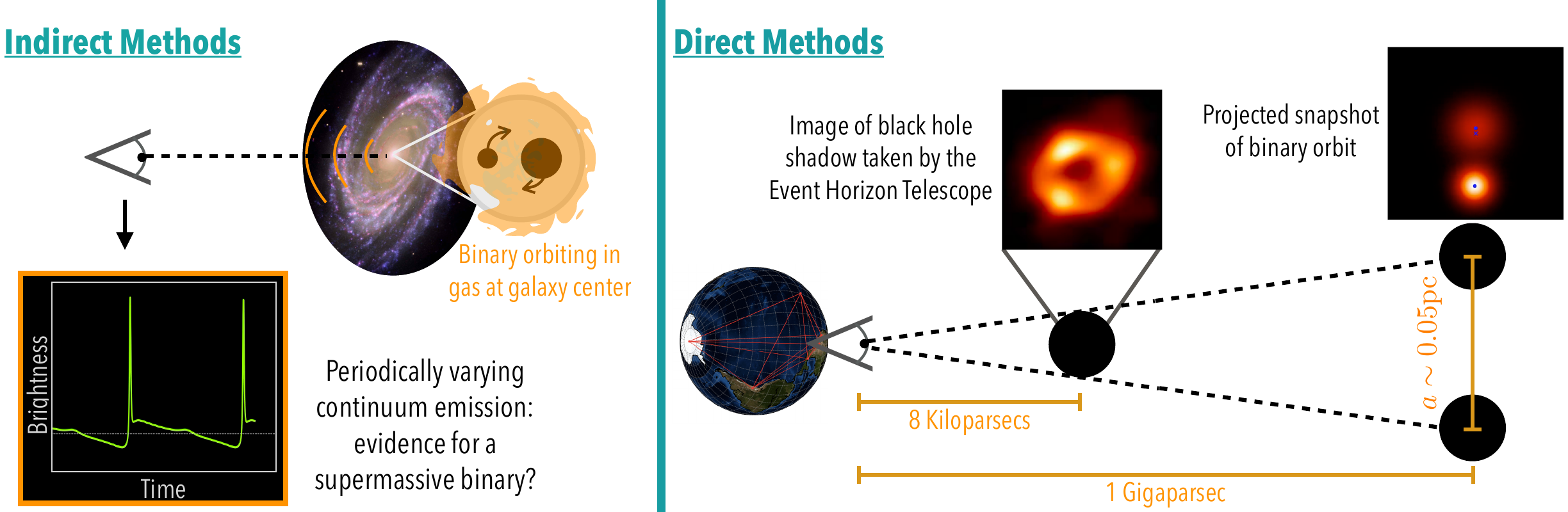} 
\end{array}$
\caption{
General classes of methods for electromagnetic identification of SMBHBs at sub-parsec separations. {\em Left}: Due to the small angular scales of SMBHBs, indirect methods for their detection are most often employed (\S's \ref{S:SpecSigs}-\ref{S:JetMorph}) {\em Right:} Only the most extreme resolving power available via, \eg, mm-wavelength VLBI, could access the required spatial scales. Gaps in our knowledge of mm-wavelength emission form SMBHBs and uncertainty in the binary population remain obstacles towards direct tracking of sub-parsec separation SMBHB orbits.
}
\label{Fig:Ind_vs_Dir}
\end{center}
\end{figure}

\noindent
Millimeter-wavelength Very Long Baseline Interferometry (mm-VLBI) can achieve $\sim10 \mu$as resolution, as evidenced by the Event Horizon Telescope, which imaged the accretion flow on horizon scales around M87 \citep{EHT_M87_I:2019} and Sgr A* \citep{EHT_SgrA_I:2022}. 
As illustrated in Figure~\ref{Fig:Ind_vs_Dir}, the angular scale of the horizon of the Milky Way SMBH (Sgr A*) is approximately the same as the angular scale of an SMBHB orbit with $0.05$~pc separation at a Gpc ($\sim 10 \mu$as). Furthermore, such a system would have an orbital period of less than $20$ years for total binary masses $M\geq3 \times 10^9 \Msun$. For a system at $500$~Mpc ($z\approx0.14$), $10 \mu$as corresponds to a binary separation of $\approx 0.024$~pc for which the binary would have an orbital period less than $20$ years for total binary masses $M\geq3 \times 10^8 \Msun$. Using a simple gas and GW-driven binary population model, limited by GWB upper limits from PTAs, and tied to the radio-loud quasar luminosity function (see \S\ref{S:pop_models}), \citet{DOrazioLoebVLBI:2018} estimate that a few to $\sim100$ of such resolvable systems are bright enough, have short enough orbital period, and large enough angular scale to track with EHT-like observations, though model uncertainties are large.

VLBI is not amenable to surveys, tracking of binary orbits would require monitoring target binary candidates identified through other means discussed above, or by GW observations with PTAs \citep{DOrazioLoebVLBI:2018}, if the host galaxy can be pinpointed within the large GW error volumes \citep{{2018MNRAS.477.5447G,2019MNRAS.485..248G}}. Assuming the binary appears as two point sources on the sky, each observation would require clusters of observations at different baseline orientations with respect to the line joining the point sources. In order to confirm the orbital motion, and therefore strong evidence for binarity, these clusters of observations would need to be repeated over the span of an orbital period.
Hence, high probability candidate systems that are bright at mm-wavelengths are required. Additionally, further modeling of the mm-emission region surrounding accreting SMBHBs and modeling of the interferometric signature is required to make further progress.

We note that the SMBHB model for the repeating flares in quasar OJ 287 posits an apocenter binary separation of $27 \mu$as. This in addition to OJ 287 being one of the brightest mm-wavelength sources on the sky make it a prime candidate for such monitoring with mm-VLBI. It is, however, unlikely that mm-wavelength emission is coming from both the secondary and the primary, as opposed to emanating from the base of the jet around the primary \citep{MOMO_OJ287+2022}. Hence, even when resolved, evidence for the binary hypothesis may be sparse, and requires a more sophisticated accretion model to interpret.

In addition to mm-VLBI, near IR, optical interferometry with GRAVITY+ could resolve illuminated broad-line or dusty regions around sub-parsec separation SMBHBs \citep{Dexter+2020}. Another exciting prospect is the possibility of $\mu$as-level astrometry for nearby SMBHBs with the ngVLA \citep{WrobelLazio:2021, WrobelLazio:2022} or {\em Gaia} \citep[][though the number of expected targets is marginal]{DOrazioLoeb_Gaia:2019} to vet SMBHB candidates.

 \begin{center}
 \begin{tcolorbox}[width=0.98\textwidth,colback={white},title={{\bf Pros and Cons: Direct, Sub-pc Orbital Tracking}}, colbacktitle=gray,coltitle=black]    
\textbf{Pros:}
\begin{itemize}
     \item Would deliver direct evidence for a sub-pc separation SMBHB on orbital timescales.
 \end{itemize}
 \textbf{Cons:}
 \begin{itemize}
     \item Requires high probability target systems and expensive, multi-epoch VLBI (or GRAVITY+ interferometric) observations  and thus would only be possible for a limited number of targets.
     \item Models for emission at wavelengths where imaging or astrometric tracking is possible are still under development.
 \end{itemize}
 \end{tcolorbox} 
 \end{center}

\subsection{\large  \large  Gravitational Waves and Multi-Messenger Constraints}
\label{SubS:GWs}
\begin{figure}[ht]
\begin{center}
\hspace{-10pt}
\includegraphics[width=\textwidth]{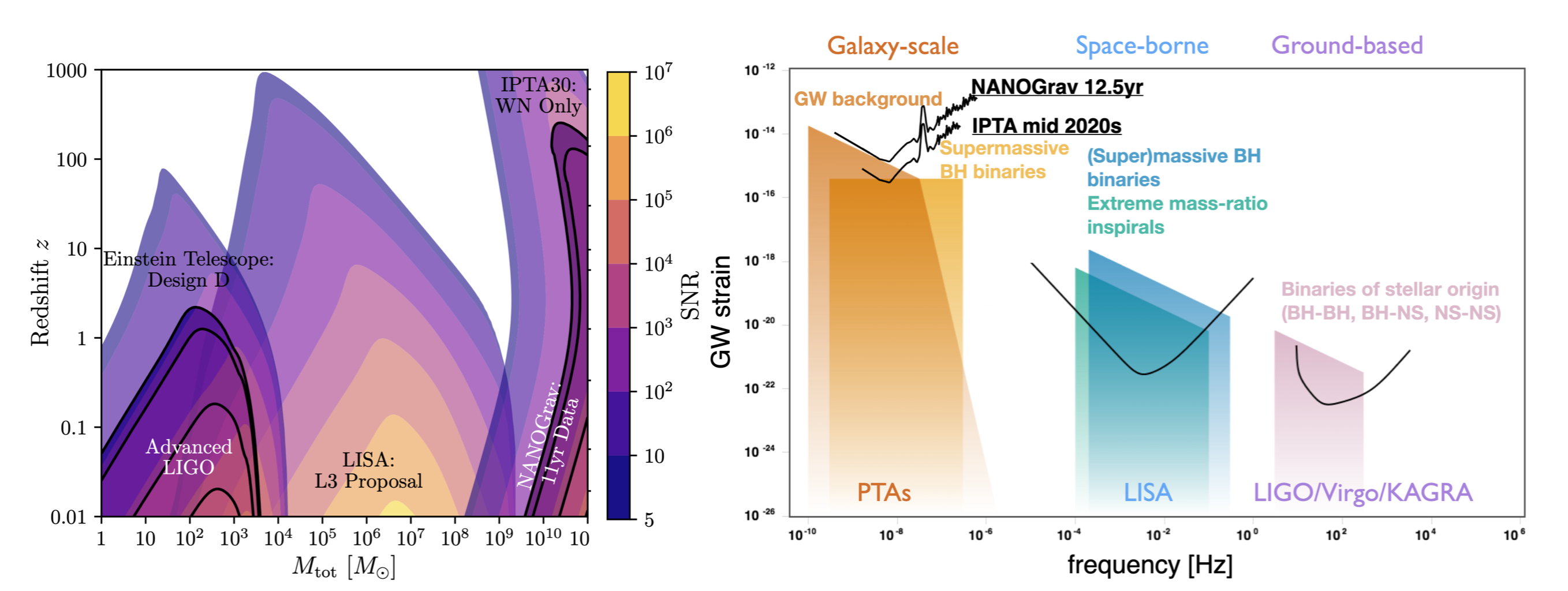} 
\caption{\emph{Left}: Signal-to-noise ratio of binaries across GW detectors as a function of total mass and redshift, adapted from \citet{KaiserMcWillliams+2021} \emph{Right}: Overview of the GW spectrum, detectors and source populations as a function of GW frequency, adapted from \citet{Moore_2015}.
}
\label{Fig:GW_Spec}
\end{center}
\end{figure}

While we have focused on electromagnetic signatures of SMBHBs, GW observations will provide a definitive handle on the population near merger, providing a boundary condition for a continuity-equation description of the population (\S\ref{S:pop_models}). The main classes of GW observatories that will observe SMBHBs are:
\begin{itemize}
    \item \textbf{Space-borne interferometers} such as the Laser Interferometer Space Antenna (LISA; \citealt{LISA:2017}) or Tian-Qin \citep{TianQin:2016} will observe in the millihertz (mHz) frequency band, $f_{\GW} \approx 10^{-4} - 10^{-1}$~Hz, and capture the mergers of $\approx 10^5-10^7 \Msun$ SMBHBs out to redshifts of $z\sim 20$ (see Figure \ref{Fig:GW_Spec} and \citealt{LISAWP:2023}). SMBHB population models place the expected LISA merger rate between of order one to several hundred per year (see below).
    %
    \item \textbf{Pulsar Timing Arrays (PTAs)} are sensitive to GWs in the nanohertz frequency band, $f_{\GW} \approx 10^{-9} - 10^{-7}$~Hz, emitted during the late inspiral the most massive $\approx 10^8-10^{10} \Msun$ SMBHBs. PTAs have recently detected evidence for a GW background \citep{2023ApJ...951L...8A,2023arXiv230616214A,2023ApJ...951L...6R,2023RAA....23g5024X}, likely produced by binaries across the universe, but dominated by SMBHBs with $z<1$ and mass $\gtrsim 10^9 \Msun$ \citep{2023ApJ...952L..37A,2023arXiv230616227A}. Individual SMBHBs in the relatively nearby universe, $z\lesssim0.5$ are expected to be resolved from the background in the near-future (see Figure \ref{Fig:GW_Spec}).
    \item \textbf{Other GW detectors} include Doppler ranging to spacecraft in the solar system \citep{Armstrong:2006}, however projections based on populations and future spacecraft missions to the Ice-Giant planets suggest that a factor of $\sim100$ improvement in sensitivity is required for these experiments to detect SMBHB mergers and contribute to, \eg, source localization \citep{SoyuerGWDop+2021}. Relative astrometric tracking of a large field of stars is also promising for detecting low-frequency GWs via the quadrupolar motions imprinted on the stellar positions.
    Upon its final data release, {\em Gaia} may yield constraints of a GW background via high precision astrometric tracking of stars in the Galaxy, covering frequencies at the high end of the PTA band and higher \citep{Moore_GaiaGW+2017}. Additionally, the Roman Space Telescope may be able to probe the frequency gap between LISA and PTAs and measure the high frequency end of a gravitational wave background from SMBHBs \citep{RomanWP:GWs}.
\end{itemize}

\vspace{2cm}

\begin{center}
    \textbf{\em Space-borne Interferometers and LISA}
\end{center}

In order to extend interferometric GW detectors, like the ground-based LIGO, to the lower frequencies needed to detect mergers of $10^4-10^7\Msun$ SMBHBs, longer interferometer arm lengths are required, along with reduced Earth-related noise sources prevalent at these frequencies \citep{LISA:2017}. Hence, planned $\sim$mHz frequency GW observatories will be space-borne. Here we focus on the joint ESA+NASA mission, LISA. LISA is nominally a four year mission that will consist of three spacecrafts in Earth-trailing orbits comprising a constellation composing the vertices of an equilateral triangle with $2.5\times10^6$~km length arms \citep{LISA:2017}. From 2015-2017, the LISA Pathfinder tested key technologies for the mission, outperforming required specifications \citep[\eg,][]{LFP_newresults+2018}, and leading the way for the full LISA mission, scheduled to launch in the early-mid 2030s.

LISA will be sensitive to a variety of GW sources including BH mergers with total masses ranging from $100\Msun$ up to $10^7 \Msun$. On overview of these sources and LISA science topics is provided in \citet{LISAWP:2023}. Here we briefly review the expected LISA merger rates for the SMBHBs relevant to this Chapter.

LISA merger rates are estimated by implementing choices for the seeding of SMBHs in galaxies and then following the subsequent evolution of the SMBHB pairs, \ie, modeling the processes described in \S\ref{S:IIphys_procs}. This is done through a combination of semi-analytical treatments \citep[\eg,][]{RicarteNatarajan_LISA:2018, Dayal_LISA+2019, Barausse_LISA+2020} and sub-grid prescriptions paired with full hydrodynamical cosmological simulations \citep[\eg, ][]{Salcido_LISA+2016, Tremmel_LISA+2018, Volonteri_LISA+2020, Katz_LISA+2020} -- a detailed overview is given in \S2.4 of \cite{LISAWP:2023}. These studies predict merger rates ranging form $\sim1$ up to a few hundred per year out to maximum redshifts ranging from $z\sim1$ to $z\sim20$.

The uncertainties in these estimates are highly dependent on seed masses of the SMBHs at birth, as well as on choices for SMBH growth through accretion over time, the effects of feedback onto the surrounding environment, the dynamics of the binary evolution and inspiral \citep[\eg,][]{Bonetti_TripIV+2019}, as well as a number of other more subtle choices that could affect, \eg, the retention of SMBHs after merger due to GW recoil \citep[\eg,][]{Villalba_LISA_recoil+2020}.

While sensitive dependence on physical parameters in models means that large uncertainties arise in predicted merger rates (and merger demographics), it also means that a measurement of the merger rate by LISA could constrain some aspects of these models. In addition to measuring merger rates, LISA will follow SMBHB inspirals for days to months before merger, observing possibly up to $\sim10^5$ cycles \citep{LISA:2017}. Long durations in band (compared to, \eg, sub-second, $\sim10$'s of cycles detections with the current ground based detectors), coupled with the LISA sensitivity, results in high signal-to-noise ratio detections for SMBHBs (Figure \ref{Fig:GW_Spec}). 
This means that LISA will be capable of measuring the orbital parameters of individual systems to high precision and so provide the possibility of providing distributions of merging binary  parameters that may be compared with predictions from formation and evolution models. This is analogous to the current efforts to understand the observed stellar-mass binary-BH mergers \citep{LIGO_03_Catalogue+2021} from individual and population masses, spins, or eccentricities. LISA could make more precise system measurements (given the much higher predicted signal-to-noise ratios), compared to current capabilities of ground-based detectors, though likely for fewer systems (given the expected rates).

\begin{center}
    \textbf{\em {Pulsar Timing Arrays (PTAs)}}
\end{center}

A PTA is a galactic scale GW detector that relies on precise measurements of milli-second pulsars \citep{2021arXiv210513270T}. These magnetized neutron stars have collimated emission along the magnetic axis, and are very stable rotators. If the rotation axis is misaligned with the radiation axis, the radiation can periodically sweep our our line-of-sight, allowing us to observe a series of precisely repeating radio pulses. If GWs from SMBHBs pass through the Galaxy, they perturb the Earth-pulsar distance, and induce coherent deviations in the arrival time of these radio signals. The GWs affect many pulsars and deviations should be detectable in many pulsars in the array. They should also be spatially correlated with a characteristic quadrupolar pattern, known as the Hellings \& Downs curve \citep{HellingsDowns:1983}. 

Currently, three main collaborations, the North American Nanohertz Observatory for Gravitational waves (NANOGrav; \citealt{NANOGrav:2013}), the European Pulsar Timing Array (EPTA; \citealt{EPTA:2013}) and the Parkes Pulsar Timing Array (PPTA; \citealt{PPTA:2013}), are systematically monitoring milli-second pulsars with the aim to detect low-frequency GWs. These, along with newer PTA collaborations, like the Indian Pulsar Timing Array (InPTA; \citealt{InPTA}), the Chinese Pulsar Timing Array (CPTA; \citealt{CPTA}) and the MeerKAT Pulsar Timing Array \citep{MeerKAT_PTA} work together under the auspices of an international consortium, the International Pulsar Timing Array (IPTA; \citealt{IPTA}).

All major PTA collaborations have recently reported evidence for a stochastic GW background with a significance between 3 and 4 $\sigma$ for the characteristic Hellings \& Downs interpulsar correlations  \citep{2023ApJ...951L...8A,2023arXiv230616214A,2023ApJ...951L...6R,2023RAA....23g5024X}. The results are broadly consistent between the different PTAs, despite differences in the datasets and data analysis methods \citep{2023arXiv230900693T}. Even though the origin of the signal is still uncertain, the background (or at least a fraction of it) is likely produced by SMBHBs throughout the universe, and can be successfully modeled with an astrophysically motivated population of SMBHBs \citep{2023ApJ...952L..37A, 2023arXiv230616227A}. The amplitude of the GW background can constrain the merger rate of massive galaxies, the SMBH mass function, and the rate of binary formation in post-merger galaxies. The shape of the GW spectrum depends on the highly uncertain orbital evolution of SMBHBs, along with orbital properties of the population of binaries, \eg, circular versus eccentric binary orbits \citep{Taylor_WP+2019, Burke-Spolaor_AstronHz+2019}.

The relatively high recovered amplitude (at the high end of theoretical expectations, \eg, see Fig.~A1 in \citealt{2023ApJ...952L..37A}) suggests efficient mergers occurring in high-mass systems. In particular, sampling the binary population parameter space with Markov Chain Monte Carlo methods to fit the GW background spectrum favors galaxy and SMBH mass functions that are towards the higher end of previous expectations (although these two parameters are degenerate) and disfavors long binary lifetimes spanning from the large separations to the PTA band. Moreover, the background spectrum shows a turnover at the lowest frequencies indicative of binary evolution driven by environmental interactions.
We note, however, that we are still in the low-signal-to-noise ratio (S/N) regime regarding the GW background and thus the population inference is sensitive to choices of the GW data analysis. Therefore, some of the above conclusions may change in the future, as the S/N increases and we are able to better characterize the background. We emphasize that, despite significant uncertainties, the GW background is the best observational evidence so far that SMBHBs form and evolve into the GW regime and eventually coalesce.

Theoretical models predict that signals from individually resolved sources can be detected on top of this background in just a few years \citep{Rosado+2015,Mingarelli+2017,Kelley_singlesource+2018, Taylor+2020}. Individually resolved signals produced by massive and relatively nearby binaries will open up the possibility for multi-messenger observations combining electromagnetic and GW data (\eg,  \citealt{Kelley_MMPTA_WP+2019,Charisi+2022}, but see also \S\ref{subS:MMA}). Since PTAs detect the early stages of GW inspiral thousands of years before coalescence, these binaries evolve slowly resulting in nearly monochromatic GW signals (over the observed baseline of observations), with long-lived electromagnetic counterparts, likely the SMBHB signatures described above. Therefore, the detection of individually resolved SMBHBs and the subsequent multi-messenger observations will allow us to probe in detail models of binary emission. Note that the slow evolution of binaries over thousands of years may be detectable via the pulsar terms, which trace the GW signal at the location of the pulsars (at kpc distances) \citep[\eg,][]{CorbinCornish:2010, McGrath_Fresnel:2021}. Already PTA limits on individually resolved binaries have provided important astrophysical constraints \citep{NANOGrav_IndvSource_Limits:2023,Falxa2023,2023ApJ...951L..50A,2023arXiv230616226A} including stringent mass ratio upper limits on galaxies in the nearby universe \citep{NANOGrav11yrGalaxies:2021}, which for some galaxies are comparable to the limits we can place on a binary companion to Sgr A* \citep{MW_companion:2023}.

PTAs have recently opened a new window to the GW spectrum finding evidence for a stochastic GW background likely produced by a population of SMBHBs. The detection of individually resolved binaries should follow soon after. Finally, PTAs in the future could explore the spatial anisotropy of the GW background \citep{Ravi+2012,Mingarelli+2013, SatoKamionkowski_aniso:2023, GardinerKelley_aniso:2023, NANOGrav_aniso:2023}, measure cosmological parameters \citep{ DOrazioLoeb_PTAH0:2021, McGrath_Hubble+2022, Wang_PTAcosmo+2022}, probe alternative theories of gravity, and constrain the density of cosmic strings and more \citep{Burke-Spolaor_AstronHz+2019}.

\begin{center}
 \begin{tcolorbox}[width=0.98\textwidth,colback={white},title={{\bf Pros and Cons: Gravitational Waves}}, colbacktitle=gray,coltitle=black]    
 \textbf{Pros:}
 \begin{itemize}
     \item GW detection is a direct method.
     \vspace{-5pt}
    \item PTAs have recently detected evidence for the stochastic GW background, consistent with production by low-frequency GWs from inspiralling SMBHBs.
         \vspace{-5pt}
    \item GW detections with PTAs and LISA will provide definitive measurements of SMBHB merger rates, which will anchor the binary formation and evolution models.
 \end{itemize}
 \textbf{Cons:}
 \begin{itemize}
\item The rates for individually resolved binaries in PTAs and LISA are highly uncertain.
     \vspace{-5pt}
    \item Even in the most optimistic scenarios, PTAs are expected to detect only a handful of binaries in the relatively near future.
     \vspace{-5pt}
    \item LISA is at least 10 years away.
     \vspace{-5pt}
    \item GWs access only the end states of the SMBHB evolution, though this reaches up to $\sim0.01-0.1$~pc separations for the most massive SMBHBs constrained by PTAs. 
%
 \end{itemize}
 \end{tcolorbox} 
 \end{center}

\section{\large  Future prospects}
\label{S:IVFuture}
We now highlight and expand upon a few future directions that we find particularly interesting.

\subsection{\large  \large  Electromagnetic Time-domain Surveys} 
\begin{wrapfigure}{r}{0.5\textwidth} 
\begin{center}
\vspace{-20pt}
\hspace{-5pt}
\includegraphics[scale=0.25]{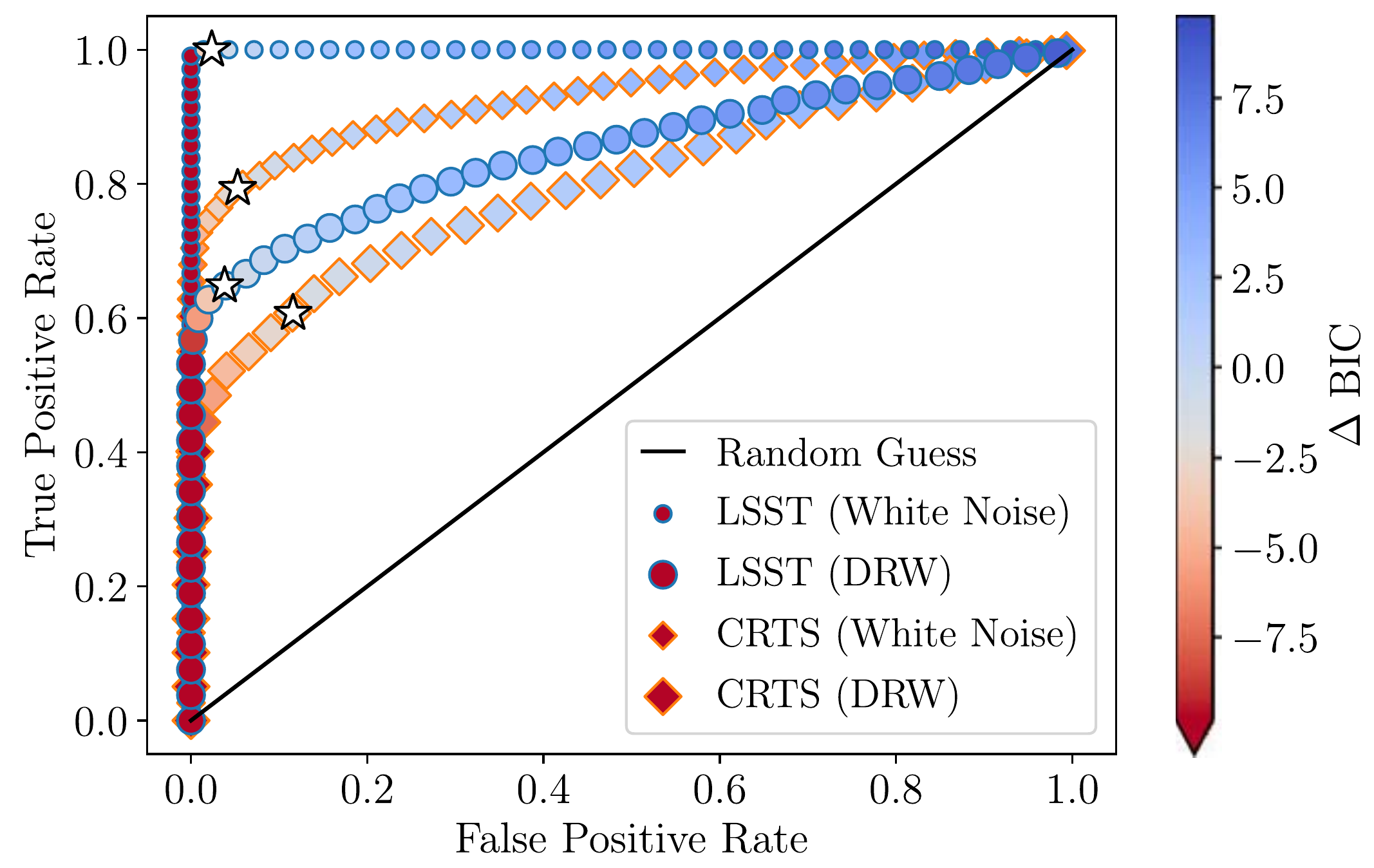} 
\caption{
True positive vs. False positive rate for detection of periodic signatures in noisy data, considering both white and DRW noise. There is an obvious improvement with LSST vs. CRTS \citep{CRTS4:Djorgovski:2011}. From \citet{WittCharisi+2022}. 
}
\label{Fig:WittLSST}
\end{center}
\vspace{-10pt}
\end{wrapfigure}

LSST with its unprecedented quantity and quality of data will revolutionize the search for SMBHBs. It will deliver a massive sample of millions of quasars, enabling the detection of rare binaries, while the exceptional sampling of the lightcurves will lead to minimal false detections (see Figure~\ref{Fig:WittLSST}).

The upcoming LSST will offer a major improvement in the observations of quasars in both survey area and depth. Early estimates of the LSST quasar population ranged from 8.4 million, for the limiting magnitude of $m=24$, achieved in a single-night LSST visit with two exposures, to 16.7 million at the full survey depth of $m=26$ achieved with co-addition of multiple epochs \citep{LSST:2019}. Revised estimates based on a recently updated luminosity function \citep{Kulkarni_QLF+2019} predict about six times more quasars ranging from 19-100 million quasars at $m=24$ and $m=26$, respectively \citep{XinHaiman_LSSTshort:2021}. The single-night limit is important for the purposes of periodicity detection, since the higher depth is achieved through co-adding of multiple images and does not allow for variability studies.

LSST will provide a large sample of quasars, in which exceedingly rare short-period binaries with periods of a few days to a few weeks will be abundant. In fact, such short-period binaries will be the sweet spot for discovery in LSST, since many cycles will repeat within the data, even with the baselines from the first few data releases. For instance, \cite{XinHaiman_LSSTshort:2021} estimated that 30-150 ultra-short period binaries (with observed period $P<1$~day) should be detectable in LSST. Since binaries spend more time at wider separations (and longer orbital periods), they predicted the detection of order $10^3 - 10^4$ binaries with periods 30-100 days considering the nightly flux limit ($m=24$). Additional theoretical models predict that a few hundred binaries should be detectable through the self-lensing signature \citep{DODi:2018} or due to relativistic Doppler boost \citep{KelleyDop+2019, KelleyLens+2021}.

Another major advantage of LSST will be its dense temporal coverage. Most of the survey time will be spent on the deep-wide-fast survey mode, which will cover the 18,000 deg$^2$ footprint with a semi-regular cadence, with repeat observations every three to five nights. This strategy will return high-quality lightcurves with dense sampling, ideal for periodicity detection. In addition, observations will alternate among six narrow-band optical filters providing invaluable color information (though making the cadence in a single color lower than once per three-five nights). \citet{WittCharisi+2022} demonstrated that the quality of the LSST lightcurves is truly a game changer in periodicity searches returning a minimal contamination with false detections; the false positive rate is significantly reduced compared to CRTS (see Figure~\ref{Fig:WittLSST}). 

The High Latitude Time Domain Survey of the planned Roman Space Telescope will cover a smaller area of the sky ($\sim 19 \deg^2$) but will go much deeper than LSST, reaching $m=28$ in a single epoch, or $m=30$ co-added. \citet{RomanPLCs+2023} predict that a four month survey would detect $\sim100$ periodic quasars. Requiring observations of 10 cycles, this allows maximum periods of $\sim12$~days, while rarity of shorter period SMBHBs and survey cadence limit to $\sim5$~days for the shortest detectable periods. The depth of a Roman survey would uniquely allow discovery of systems powered by SMBHBs in the $10^{4-6}\Msun$ range, which will merge in the LISA band after $10^{3-5}$~yrs \citep{RomanPLCs+2023}. 

We also note that an X-ray Survey mission \citep[the proposed Star-X mission, \eg,][]{StarX:2022} would benefit periodicity searches, especially for self-lensing signatures where emission coming from a small region near the SMBH can be the most magnified.

\subsection{\large  \large  Direct Orbital Tracking with High-angular Resolution Experiments}%

As discussed in \S\ref{S:DirImg}, present and next-generation VLBI experiments and the GRAVITY+ optical interferometer will be able to access $\mu$as scales subtended by sub-parsec separation SMBHBs, via imaging and astrometric photo-center tracking. While difficult, the astrometric or resolved tracking of an SMBHB orbit over the course of a period offers what many other methods do not, an unambiguous electromagnetic identification of a sub-parsec separation SMBHB. This definitive detection could help us to  build a detection ladder by studying the properties of the system on all scales and wavelengths and to possibly use this information to devise new search methods. With enough definitively identified SMBHB systems, we may be able to start characterizing how SMBHB-harboring galaxies differ in their properties, if at all, from the ``normal quasars.''

\subsection{\large  \large  Low-frequency Gravitational Wave Experiments \& Multi-messenger Promise}
\label{subS:MMA}
As mentioned above, PTAs have already reached a milestone by finding evidence for the low-frequency GW background in the nanohertz band, the significance of which is expected to grow in the upcoming years leading to a definitive detection. Upper limits on the GW background had already provided significant constraints onto the population of binary candidates detected in time-domain surveys \citep{Sesana+PLC_PTA+2018,Holgado+2018}. These multi-messenger consistency checks on close-separation SMBHB candidates will become even more powerful in vetting the candidate population, as the PTA sensitivity increases and the background is measured across the PTA frequency range. Several factors contribute to the expected PTA sensitivity increase, such as: (1) the pulsars are monitored for longer baselines enabling us to probe lower frequencies of the GW spectrum, (2) technological advances in the radio receivers reduce the noise of observations, (3) more advanced statistical techniques are employed for the analysis of the data, (4) more and higher quality pulsars are discovered and are systematically monitored. The combination of multiple datasets, e.g., from individual PTAs into a common International PTA dataset can also boost the sensitivity. In addition, existing radio telescopes like the Canadian Hydrogen Intensity Mapping Experiment (CHIME; \citealt{CHIME}) and the Five-hundred-meter Aperture Spherical Telescope (FAST; \citealt{FAST}), along with future radio observatories, like ngVLA \citep{BurkeSpolaor_nGVLA_SMBHBs:2018}, the Deep Synoptic Array (DSA-2000; \citealt{DSA_2000}), and the Square Kilometer Array (SKA; \citealt{SKA_PTA}) are expected to vastly increase the number of detected and monitored pulsars, greatly improving the sensitivity of PTAs.

\begin{figure}[h]
\begin{center}$
\begin{array}{c}
\hspace{-10pt}
\end{array}$
\includegraphics[scale=0.5]{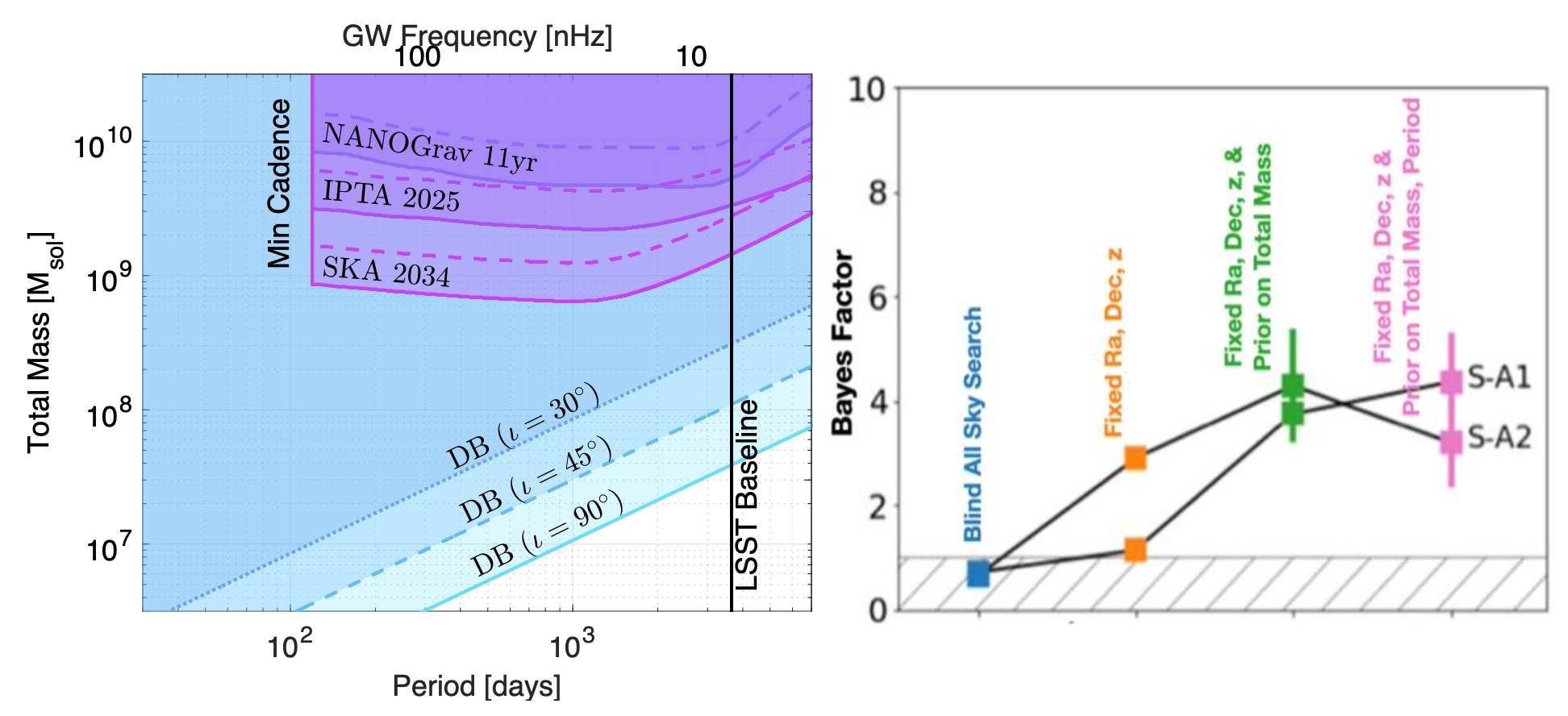} 
\caption{\emph{Left}:
Binary parameter space for which multi-messenger observations with time-domain surveys and PTAs are possible. The blue shaded regions show EM detectable binaries with peak-to-peak Doppler boost variability $\geq10\%$
for different inclinations and fixed mass ratio of $q=1/4$. The purple shaded regions delineate the GW detectability of binaries of the same mass ratio with current and projected PTA sensitivity. From \citet{Charisi+2022} \emph{Right}: Comparison of Bayes factors for GW detection between completely uninformed all-sky searches and searches with some prior information about the host galaxy or the binary. Adapted from \cite{LiuVig_I:2021}.
}
\label{Fig:multi-messenger}
\end{center}
\end{figure}

Given the discovery of evidence for the GW background, PTAs are also expected to detect GWs from individual SMBHBs that produce signals strong enough to stand above the background \citep{Rosado+2015,Mingarelli+2017,Kelley_singlesource+2018, Taylor+2020}.
Since SMBHBs are expected to emit both bright electromagnetic signals and strong GWs, they make exceptional targets for multi-messenger observations \citep{Sesana_MMPTAXray:2012, Kelley_MMPTA_WP+2019, Baker_MMmHz_WP+2019, Charisi+2022}. 
Multi-messenger observations will offer significant advantages in constraining the dynamical interaction of binaries with their environment, unveiling the physics of accretion in the presence of binaries, and probing the co-evolution of SMBHs with their host galaxies, and more \citep{Kelley_MMPTA_WP+2019}. PTAs and current electromagnetic time-domain surveys probe the overlapping populations of SMBHBs, i.e. binaries with similar periods and masses (see Figure~\ref{Fig:BHBDemo}, and \citealt{Charisi+2022}). In Figure~\ref{Fig:multi-messenger}, we show the parameter space of SMBHBs, where multi-messenger observations are possible given current and future PTA capabilities.
Combining electromagnetic and PTA data has great advantages; incorporating binary orbital constraints from electromagnetic observations (e.g., period/mass of a candidate) can improve GW upper limits by a factor of two \citep{NANOGrav_MMsearch_X2:2020}.

Similarly, including priors from electromagnetic data in PTA searches can boost the detectability of binaries and improve parameter estimation \citep{LiuVig_I:2021, LiuVig_II:2023, charisi_inprep}. In the right panel of Figure~\ref{Fig:multi-messenger}, we show the Bayes factor for GW detection for simulated binaries injected in a PTA that resembles the NANOGrav 11yr dataset. The detectability of the same binary increases when prior information \eg, known from an electromagnetic candidate is included in the search. In the case of the Doppler boost (and self-lensing) variability, the PTA signal can be directly linked to the periodicity of a quasar candidate for a multi-messenger data stream \citep{Charisi+2022}, potentially allowing for further boost in the detectability of the source and its parameter estimation.

In the LISA-era of the next decade, electromagnetic counterparts due to a plethora of sources have been and continue to be explored, including circumbinary disk and mini-disk interactions, Doppler boosting and jet launching in the moments before and through merger \citep{Palenzuela_dbljet+2010, Noble+2012, Moesta_subdom_BHBjets2012, Gold:GRMHD_CBD:2014, Gold:GRMHD_CBDII:2014, Tang_LateInspCool+2018, Farris:2015:GW, Kelly_LISAEMctpt+2017, Haiman:2017, Kahn+2018, dAscoli+2018, Kelly_LISAEMctpt_II+2021, Gutierrez+2022, BrightPasch_minidisks:2023, WangSongsheng+2023} as well as post-merger GW recoil induced shocks \citep{Lippai_Z3+2008, Corrales+2010, Rossi+2010, Meliani_recoil+2017, Krauth+2023}. If one of these electromagnetic signatures arises, we can only observe it if we know in advance which direction on the sky to point telescopes. \citet{Mangiagli+2020} considers LISA's ability to localize a GW source on the sky in time to follow up with electromagnetic observations and follow up work \citep{Mangiagli+2022} considers optical, radio, and X-ray counterparts to SMBHB mergers and determines detectability with joint LSST plus LISA, the Square Kilometer Array \citep[SKA][]{SKA:2009} plus the European Extremely Large Telescope \citep[][]{ELT} plus LISA, or Athena \citep{ATHENA:2013} plus E-ELT plus LISA configurations. They determine that a few such electromagnetic counterparts to SMBHB mergers may be detectable during a four year LISA mission. Further possibilities with X-ray and GW joint detections with a LISA and Athena are detailed in \citet{Piro_LISA_ATHENA+2022}.

\section{\large  Summary of observational searches and candidates}
Below we summarize the observational efforts to uncover SMBHB candidates based on the signatures described in this Chapter.

\label{Sec:obs_summary}
\subsection{\large  Searches and candidates from spectroscopic signatures}
\label{apendix:Spec}
\subsubsection{\large  Systematic Searches for Broad Emission Line Shifts in Optical Spectra from SDSS}
\begin{itemize}
\item \cite{Tsalmantza+2011}: Search for broad lines offset with respect to narrow lines in $\sim$59,000 quasars with $0.1<z<1.5$ revealed 34 sources with peculiar spectra, of which 9 are binary candidates (4 new, and 5 previously known).
\vspace{-5pt}
\item \cite{Erac+2012}: Search for displaced broad $H\beta$ lines in 15,900 quasars with $z<0.7$ found 88 candidates (some previously known). Of those, 68 candidates were followed-up revealing significant velocity shifts in 14.
\vspace{-5pt}
\item \cite{Ju+2013}: Search for velocity shifts in broad Mg II lines in $>1500$ quasars with $0.36<z<2$ and multiple spectra detected 7 (64) strong (possible) candidates. They inferred that $\leq18\%$ of quasars can host binaries.
\vspace{-5pt}
\item \cite{Shen_BLsearchI+2013}: Search for broad $H\beta$ line shifts in 700 quasars with  $z<0.7$ and multiple spectra revealed 28 sources, of which 7 are top binary candidates. If variability is due to binaries and not intrinsic, most quasars should harbor SMBHBs.
\vspace{-5pt}
\item \cite{LiuShen_BLsearchII+2014}: Search for displaced $H\beta$ broad lines in $\sim$ 21,000 quasars found 399 candidates, of which 50 were followed-up revealing 24 sources with significant radial accelerations and 9 were considered top binary candidates. Unlike previous studies \eg, \cite{Erac+2012}, moderate velocity offsets ($<$1000km/s) were also considered.
\end{itemize}

\subsubsection{\large  Spectroscopic Follow-up of Candidates Identified in Systematic Searches}
\begin{itemize}
\item \cite{Decarli+2013}: Follow-up of the \cite{Tsalmantza+2011} sample found significant line variations in 11 sources. It is unclear whether the variability is coherent as expected for binaries or transient and longer monitoring is required. 
\vspace{-5pt}
\item \cite{Runnoe+2015, Runnoe+2017}: Follow-up of the \cite{Erac+2012} sample over 12 years with 3-4 observations per source. Radial velocity curves can be constructed for 29 sources with 3 candidates showing monotonic velocity changes. Limits on binary parameters cannot exclude any of the candidates, but this will change with longer baselines.
\vspace{-5pt}
\item \cite{Wang+2017}: Follow-up of 21 candidates of the \cite{Ju+2013} sample found that none is consistent with the binary scenario. The search was repeated in a sample $\sim$1400 quasars with multi-epoch spectra spanning 10 years from BOSS and revealed one candidate. They concluded that $\leq1\%$ of the quasars host binaries with separation of order $\sim0.1$~pc.
\vspace{-5pt}
\item \cite{GuoShen_BLsearchIII+2019}: Follow-up of 52 candidates in \cite{Shen_BLsearchI+2013, LiuShen_BLsearchII+2014} showed that 5 sources are viable candidates based on their radial velocity curves, concluding that $\sim$13\% of quasars are associated with sub-parsec binaries.
\end{itemize}

\subsubsection{\large  Individual Candidates Primarily from Early SDSS Studies}
\begin{itemize}
\item J0927+2943: Highly blue-shifted Balmer lines could indicate a recoiling SMBH \citep{Komossa+2008recoil} or an SMBHB \citep{Dotti+2009,Bogdanovic2009candidate}, but both hypotheses were later disfavored by \cite{Decarli2014}.
\vspace{-5pt}
\item J1536+0441: Candidate identified by atypical spectrum with two broad emission line systems \citep{BorosonLauer:2009}.
\vspace{-5pt}
\item J1050+3456: Candidate with blue-shifted Balmer lines, but could be a double-peaked emitter \citep{Shields+2009}.
\vspace{-5pt}
\item J1000+2233: A binary was invoked to explain extreme velocity offsets in $H\alpha$ and $H\beta$ lines \citep{Decarli2010}.
\vspace{-5pt}
\item J0932+0318: Similar to J1050+3456, double-peaked emission is a possible alternative to binarity \citep{Barrows2011}.
\vspace{-5pt}
\item NGC4151: Lightcurve and radial velocity curve could be explained by an eccentric binary candidate \citep{Bon+2012}.
\vspace{-5pt}
\item Mrk 231: The observed flux deficit in the UV/optical spectrum was associated with a cavity in a circumbinary disk \citep{Yan_Mrk231_notch+2015}, but dust reddening can better explain the entire spectrum \citep{Leighly+Mrk231+2016}. 
\end{itemize}

\subsubsection{\large  Candidates with Double-peaked Spectra and Follow-ups}
\begin{itemize}
\item Quasars with double-peaked broad emission lines, like QSO OX 169, Arp 102B, 3C 390.3, 3C 332 were proposed as binary candidates \citep{Gaskell:1983,StockFarn:_OX169:1991,Gaskell+1996}. 
\vspace{-5pt}
\item Subsequent follow-up monitoring of the above and many other similar sources disfavored the SMBHB interpretation for this class of quasars, known as double-peaked emitters. The line variability is inconsistent with expectations for binaries and they can be more naturally explained by disks around single-SMBHs \citep{HalpFili:1988, EracHalp:1994, Erac_CandRejec:1997, EracHalp_DblPeak:2003,Erac_Dblpeak_AGNdsk:2009, LiuErac+2016,DoanErac+2020}.

\end{itemize}

\subsection{\large  Searches and Candidates from Photometric Signatures}
\label{apendix:Phot}

\subsubsection{\large  Systematic Searches in Time-domain Surveys}
\begin{itemize}
\item \cite{Graham+2015b}: Search in a sample of 243,500 spectroscopically confirmed in CRTS revealed 
111 candidates with periods of 650-2500\,d. The lightcurves of the candidates were extended with data from LINEAR when available. 
\vspace{-5pt}
\item \cite{Charisi+2016}: Search in a sample of 35,383 quasars in PTF selected 50 candidates with periods of 130-900\,d, 33 of which remain statistically significant after the re-analysis of extended lightcurves with data from CRTS and iPTF. 
\vspace{-5pt}
\item \cite{LiuGezari+2019}: Search in a sample of 9000 quasars from Pan-STARRS MDS selected 26 candidates with periods of 300-1000\,d, but follow-up monitoring with the Discovery Channel Telescope and LCO excluded all but one candidate.
\vspace{-5pt}
\item \cite{Chen_DES_PLCs+2020}: Search in a sample of 625 quasars in overlapping DES+SDSS stripe 82 fields with long baseline lightcurves (20yr) detected 5 candidates with periods of 3-5\,yr.
\vspace{-5pt}
\item \cite{Chen_ZTF_PLCs+2022}: Search in 143,700 quasars from ZTF detected 127 candidates with periods of 500-1000\,d. 
\vspace{-5pt}
\item \cite{Swift_BAT_Serafinelli,Swift_BAT_Liu} Two independent searches in Swift/BAT 105-month survey. The former analyzed 553 quasars, found a new candidate, and validated a known one (MCG+11-11-032). The latter analyzed 941 quasars and found no evidence for periodicity including the previously known candidate.
\vspace{-5pt}
\item \cite{Sandrelli+2018, Holgado+2018}: These two studies (and references therein) summarize the binary candidates identified from their quasi-periodic behaviour in Fermi gamma-ray lightcurves.
\end{itemize}

\subsubsection{\large Periodic Candidates Identified Individually}

\begin{itemize}
\item Optical: 
    \begin{itemize}
        \item OJ 287: This bright blazar has a lightcurve spanning a century (albeit sparse initially) with quasi-periodicity of 11.86yr and twin flares 1yr apart. The periodicity is explained by an eccentric and unequal mass binary; the flares are produced every time the secondary plunges through the primary's accretion disk on a relativistically precessing orbit \citep{Valtonen_Oj287+2008,MOMO_OJ287+2021}.
        The most recently expected flare, however, was not confirmed by observations \citep{Komossoa_noOj287_flare+2023, Valtonen_noflareResponse+2023}.
        \item PG 1302-102: The first candidate found in the recent systematic searches \citep{Graham+2015b}. Because of its brightness it allowed for extensive follow-up studies, which potentially reveal additional evidence for its binary nature. 
        \item SDSS J0159+0105: This candidate was detected in CRTS, although it was not selected by \cite{Graham+2015b}. It has two periods (741 and 1500 days) and was interpreted as having the predicted by simulations 2:1 period ratio. The broad H$\beta$ line exhibits variability and it has unusually bright UV spectrum \citep{Zheng+2016}.
        \item Q J0158-4325: Periodicity of $\sim176$\,d was detected in this microlensed quasar \citep{Millon+2022}.
        \item J025214.67−002813.7: One of the 5 candidates detected in DES+SDSS search with a period of $1607$\,d was associated with accretion variability models \citep{Liao_MdotPLC_cand+2021}.
        \item Spikey: This candidate was identified from its symmetric flare \citep{SpikeyHu+2020}. The variability is well fit by an eccentric self-lensing model, but its periodicity has not been confirmed yet.
    \end{itemize}
\item Radio: 
    \begin{itemize}
        \item PKS 2131−021: This candidate has a periodicity of $1729$\,d in the 14.5-15.5 GHz lightcurves, and was first put forth by \citet{Ren+PKS0153_2021}. The periodicity disappears and appears almost in phase at a later epoch and was interpreted with a jet-Doppler model by \citet{ONiell_ReadBland+2022}.
    \end{itemize}
\item Gamma Rays: 
     \begin{itemize}
        \item PG 1553+113: The first gamma-ray variable blazar with a detected periodicity of $2.18$\, y in rest frame ($z=1.285$) \citep{gamPLC_Fermi+2015, Sandrelli+2018} has prompted many multi-wavelength follow-up studies.
        \item PKS 2247−131: A 34.5\, period was attributed to jet helicity due to an orbiting SMBHB \citep{Zhou_34p5day_gamPLC+2018}.
        \item PKS 2155−304: A periodicity of $620$\,d was detected in this candidate \citep{Sandrelli+2018}.
        \item BL Lac: Candidate identified from periodicity of $1.86$\,yr \cite{Sandrelli+2018}.
    \end{itemize}
\end{itemize}

\subsection{\large  Candidates from Jet Signatures}
\label{apendix:jets}

\subsubsection{\large  Individual Candidates with Helical or Periodic Jet Structures}
\begin{itemize}
        \item S5 1928+738: The helical structure and periodic components of the jet can be explained by the orbital motion of the jet-launching SMBH in a binary system \citep{Roos:1993, Kun+2014}.
        \vspace{-5pt}
        \item Mkn 501: A binary model could explain the complex structure of the curved  helical jet \citep{Villata1999}.
        \vspace{-5pt}
        \item 3C 273: The rapid jet precession may be due to precession of in the inner accretion disk induced by a second SMBH in a non-co-planar orbit \citep{AbrRom:1999, Romero+2000}.
        \vspace{-5pt}
        \item PKS 0420-014: A binary model can explain the motion of jet components on helical trajectories \citep{Britzen2001}.
        \vspace{-5pt}
        \item BL Lacertae (2200+420): A straight jet precessing with a short  period could be caused by an SMBHB \citep{Stirling2003}, in which the jet is associated with the secondary (less massive) SMBH \citep{Caproni2013}.
        \vspace{-5pt}
        \item 3C 120: A binary was invoked to explain the helical jet structure and optical periodicity \citep{Caproni2004}.
        \vspace{-5pt}
        \item AGN 1156+295: The oscillatory pattern in the jet could be explained by a binary candidate \citep{Hong2004}, but follow-up observations are more consistent with a Kelvin Helmholtz instability \citep{2011A&A...529A.113Z}.
        \vspace{-5pt}
        \item 3C 345: The position and velocity of multiple jet components suggest a wobbling jet with a short precession period consistent with an SMBHB \citep{Caproni2004_3c345,Lobanov2005,2009RAA.....9..137Q}.
        \vspace{-5pt}
        \item PKS 1830-211: The helical jet may be caused by rapid precession due to an SMBHB \citep{Nair2005}.
        \vspace{-5pt}
        \item S5 1803+784: A binary was proposed to explain the evolution of the VLBI coordinates of a single jet component \citep{Roland2008}, but none of the existing models can explain all the jet properties \citep{Britzen2010}.
        \vspace{-5pt}
        \item B1308+326: The kinematics of multiple jet components can be explained by jet precession due to an SMBHB, but other explanations (\eg, spin precession of a single SMBH) are also possible \citep{Qian2017}.
        \vspace{-5pt}
        \item 3C 279: The complex jet structure of this candidate is consistent with two precessing jets with the same period but different orientations \citep{Qian2019}, but detailed observations by EHT challenge this model \citep{EHT_3c279}.
        \vspace{-5pt}
        \item 3C 454.3: The kinematic behavior of multiple jet components suggests the existence of two groups of knots produced by a double-jet in a binary system \citep{Qian2021}.
        \vspace{-5pt}
        \item 4C31.61 (2201+315): The jet structure can be explained by a system of three distinct jets, associated with three SMBHs separated by $\sim1-2$\,pc \citep{Roland2020}.
           \vspace{-5pt}
        \item \citet{Britzen_JetWiggles+2023} apply jet precession models to OJ 287 as well as 11 other quasars listed in their Table 6 (3C 279, 3C 273, PKS 0735+178, 2200+420, PG 1553+113, 3C 345, 3C 120, 1308+326, TXS 0506+056, 3C 84, PKS 1502+106).
\end{itemize}

\subsubsection{\large  Follow-up of Candidates Selected with Other Methods}
\label{apendix:jets:followup}
\begin{itemize}
        \item Spikey: VLBI images show a wiggly jet structure consistent with the binary scenario \citep{Kun+2020:Spikey}.
        \vspace{-5pt}
        \item PG1302: The morphology of the jet at kilo-parsec scale along with parsec scale kinematics support the binary scenario for this candidate \citep{ Kun+2015:PG1302, Mohan_PG1302+2016}.
         \vspace{-5pt}
        \item OJ287: The highly variable jet of this source can be fit by a helical model produced by the binary orbital motion \cite{2013A&A...557A..28V}. 
          \vspace{-5pt}
        \item PG1553: The kinematics of several jet components suggest a precessing jet consistent with the periodicity detected in the gamma-ray lightcurve \citep{2017ApJ...851L..39C}.
\end{itemize}

\subsubsection{\large  X and Z-shaped Jets}

\begin{itemize}
        \item \cite{Yang+Xjet+2019} present a catalog of 290 “winged” or X-shaped radio galaxies.
\end{itemize}

\section*{Acknowledgments}
The authors would like to thank Jessie Runnoe, Aaron Stemo for comments on an early version of the manuscript. D.J.D. received funding from the European Union's Horizon 2020 research and innovation programme under Marie Sklodowska-Curie grant agreement No. 101029157, and from the Danish Independent Research Fund through Sapere Aude Starting Grant No. 121587. MC acknowledges support from NSF award AST-2007993.
\clearpage

\bibliographystyle{apj} 
\bibliography{refs}
\end{document}